\documentstyle[epsfig]{mn}
\title{A revision of the solar neighbourhood metallicity distribution}
\author[M. Haywood]
       {M. Haywood\thanks{email : Misha.Haywood@obspm.fr} \\
        DASGAL, URA CNRS   8633, Section d'Astrophysique, Observatoire de  Paris, F-92195 Meudon Cedex,France}
\date{Accepted.
      Received ;
      in original form }
\pagerange{\pageref{firstpage}--\pageref{lastpage}}
\pubyear{2001}

\begin{document}

\maketitle

\label{firstpage}

\begin{abstract}
We present a revised metallicity  distribution of dwarfs in the  solar
neighbourhood. This distribution is centred  on solar metallicity.  We
show that previous metallicity distributions, selected on the basis of
spectral type,   are biased against  stars  with solar  metallicity or
higher. A selection  of G-dwarf  stars  is  inherently biased  against
metal rich stars and is not representative  of the solar neighbourhood
metallicity distribution. 
Using a  sample selected  on colour, we  obtain a  distribution  where
approximately   half   the stars  in   the  solar neighbourhood  has a
metallicity  higher  than [Fe/H]=0.  The  percentage of mid-metal-poor
stars ([Fe/H]$<$-0.5)  is approximately 4 per  cent, in agreement with
present estimates of the thick disc. \\ 
In   order to have  a  metallicity distribution comparable to chemical
evolution model  predictions, we  convert the  star fraction  to  mass
fraction, and show that another  bias against metal-rich stars affects
dwarf metallicity distributions, due to  the colour (or spectral type)
limits of the samples. 
Reconsidering the corrections due  to the increasing thickness of  the
stellar disc with  age, we show that  the Simple Closed-Box model with
no instantaneous recycling approximation gives a reasonable fit to the
observed distribution. 
Comparisons with the  age-metallicity  relation and abundance   ratios
suggest that the Simple Closed-Box model may be a  viable model of the
chemical evolution of the Galaxy at solar radius. 
\end{abstract} 


\section{Introduction}

The aim  of  this paper is  to  present and discuss  a new metallicity
distribution of the stellar material found in the solar neighbourhood.
In recent  years,   several   such  metallicity   distributions,
constructed from various samples  of  solar neighbourhood dwarfs  have
been published  \cite{96ROC_EA3,95WYS_EA,97FLY_EA,97FAV_EA2}. 
These studies have repeatedly pointed to a deficit  of metal-poor stars
relative   to   the  simplistic  but    insightful  Simple  Closed-Box 
model (hereafter SCB model).
However, since the early seventies, models have successfully
fit the observed metallicity distribution by alleviating one or another assumption
of the SCB model
\footnote{The assumptions of the  SCB model are that the solar neighbourhood
is considered as closed box, with no matter flowing in or out,
the gas is initially free of metals, the initial mass function
is constant, and the interstellar medium is well mixed at all times.}.  
Such fits  are now routinely obtained
with present day models  of chemical evolution  of the Milky Way  \cite{97CHI_EA,98PRA_EA}  using a   pletora  of
solutions, the most widely accepted being the infall model. \\
In view of the apparent easiness of these models to fit this 
constraint of chemical evolution, one  may wonder why the deficit  of
metal-poor stars is regularly addressed, and why we intend to do so. 
One reason is that the completion of Str\"omgren surveys by Olsen 
(see Olsen (1983) and thereafter) has considerably extended the available data. 
Moreover, the coincidental publication of the $Hipparcos$ Catalogue now allows a 
clean definition of a complete sample of solar neighbourhood dwarfs.

A second reason  is that the  role of the  thick disc is still unclear
and difficult to evaluate : is the thick disc the first epoch of the disc formation ?
Is it cogenetic to the Galaxy ? And how should it be taken into account when considering
the G dwarf problem ? In the present study, we take the view that the thick disc is an 
integral part of the disc and that it should be considered when discussing the G-dwarf 
problem.
This raises the question of the exact contribution of the thick disc
at [Fe/H]$<$-0.4. Wyse \& Gilmore (1995) have contended that the important metallicity 
tail below this limit ($>$20\%) is likely due  to contamination by the thin disc.
This problem is then connected to another. If the thin disc contribution at 
[Fe/H]$<$-0.4 is as important as envisaged by Wyse \& Gilmore (1995), it leads
to the conclusion that the SFR must have been very efficient 
at early times in the disc. This conclusion is reinforced by the apparent rapid
rise of metallicity in the age-metallicity relation \shortcite{97SCU_EA} - leaving
even less time to produce the material at [Fe/H]$<$-0.4.
As a matter of fact, it is a feature common to most models with infall that they use a decreasing 
SFR, in accordance with the infall decay rate \shortcite{96TOS}.
In contrast, all recent determinations of the SFR history of the disc point to
a constant, or even slightly increasing SFR \cite{97HAY_EA2,00BIN_EA}.
We show here that these apparent contradictions are mostly an effect
of pre-Hipparcos data, and that the metallicity tail at [Fe/H]$<$-0.4 is 
drastically reduced with Hipparcos parallaxes.

Third, while chemical evolution models have increased considerably in complexity
over the last decade, the sophisticated solutions envisaged to solve 
the `G-dwarf problem' (such as infall, variable initial mass function, etc), are still
essentially adhoc, and  direct observational evidence favoring one or another alternative have 
remain extremely elusive. 
Infall models have focused most of the efforts of galactic chemical evolution studies in recent 
years, while relatively little attention has been devoted to other solutions.
Since the G-dwarf problem remains the main (indirect) support for long time-scale 
gas accretion by the Milky Way disc, it is important to check the failure (or success) of 
no-infall models. Also, when considering scenarios where the thick disc is envisaged
as a genuine galactic population (as is the case here), 
it must be born in mind that no-infall models may come as a more natural alternative.
The existence of the thick disc, if considered as cogenetic, implies that a substantial stellar disc must have been in place at early 
times (Wyse, 2000 astro-ph/0012270).
As remarked by Wyse, this suggestion could find some support or invalidation in a more general context,
from studies of high redshift galactic discs  \shortcite{00BRI_EA}.

Finally, another incentive for this work has been the (naive) realisation by the author that 
while models of chemical evolution predict distributions of
stellar {\it mass} as a function of metallicity, previous studies of the dwarf
metallicity distributions have only been able to give statistics of {\it stars}
as a function of metallicity. 
The accurate positioning of the stars in the HR diagram allowed by $Hipparcos$
parallaxes permits the conversion from (colour, magnitude) to masses.
It will be demonstrated that, while the conversion itself has a relatively minor 
effect, the colour limits of the sample bias the distribution by underestimating
the contribution of solar-metallicity and metal-rich stars.

The remainder of the paper is divided into 5 parts. 
In the following section, we present the calibrations used to estimate metallicity
from Geneva and Str\"omgren photometry. After selecting a basic sample from considerations
of completeness and available photometry, we select long-lived stars and discuss 
the biases introduced by our selection procedure. A first estimate of a corrected metallicity 
distribution of star frequency distribution is given.
This distribution is then compared to previous works in section 3, where we also
discuss the problem of biases in other samples.
We evaluate the proportion of X-ray emitter stars in the sample.
Since X-ray coronal activity is usually considered as a tracer of young stars,
we discuss the significance of the high percentage of young star candidates in the sample.
In section \ref{finaldis} we convert our dwarf metallicity distribution to a 
mass metallicity distribution.
A proper correction for the mass bias and a final metallicity distribution are given
for both iron and oxygen abundances.
In section \ref{models}, we briefly review the observed parameters of the disc 
that enter the SCB model, and following Occam's Razor, we explore 
how the Simple model fits other local constraints of chemical evolution.
We conclude in the last section.

\section{The observed dwarf metallicity distribution}
\subsection[]{Description of the sample}

The solar neighbourhood, as sampled by the Hipparcos Catalogue, 
is considered to be essentially complete to $M_v$=8.5
for stars with $\pi>$40\-~mas \cite{97JAH_EA}, and contains 959  entries within these limits.
At $M_v$=8.5, main sequence stars have $B-V$ colour between 1.3 and 1.5.  Present day 
available photometric metallicity  calibrations are limited to bluer colour.
The Geneva photometric system  is  probably the best suited
system in that  regard, since  the calibration  is available  down  to
$B_2-V_1$=0.65   \cite{78GRE},     corresponding    to   $B-V$=1.05
approximately. 
According to the available data, the main sequence of the galactic globular cluster M92 
([Fe/H]$\approx$-2) 
passes through the point B-V=1.00, M$_v$=8.0 \shortcite{88STE_EA}.
This means that the limit at M$_v$=8.5 is unlikely to produce any significant
bias against low metallicity stars in the sample.

Within these limits, the  sample contains 681 entries.  
If we further select  stars with M$_v>$3.5  and $B-V>$0.25 and which are not
flagued  as  either G,  O, S, or   V suspected  binaries, we  find 475
entries, of which 82 are flagued ``C'' component solutions in the main
catalogue.  For 27   of these 82   systems, no Geneva
photometry  is available from   the   General   Catalogue of    Photometric Data
\cite{97MER_EA1}. Those ``C'' component solutions for which photometry is available
are usually not separated or  simply not detected as binaries in the
GCPD data base (except
for 3 systems, for which photometry for the primary is available).  We
decided therefore to exclude entries  with the flag ``C'' from further
consideration, keeping in mind that we are however excluding a minimum
of 82  stars with $B-V<$1.05 from our  sample.  We  are then left with
393 stars for which iron abundance estimates are necessary.

\subsection{Photometric metallicity calibration}

When dealing with the metallicity distribution of stars in the  solar
neighbourhood, one is interested by long-lived stars down the main sequence, 
in order to avoid bias favoring young -- and possibly more metal-rich -- objets.
However, metallicity measurements and calibration of M stars are still in infancy, 
and one is limited to stars bluer than M spectral type. 
There is no one single metallicity indicator for all the stars in the samples
analysed in Sec.  3 and 4 and we rely on Str\"omgren or
Geneva  photometric metallicity, and for  a few  stars on  spectroscopic data.   We
discuss herebelow the photometric calibrations we use, by reference to
spectroscopic measurements.

\subsubsection{Geneva photometric calibration for dwarfs : Grenon (1978)}

A metallicity calibration   of Geneva colour  indices exists  for main
sequence stars  down   to $B_2-V_1$=0.65 ($B-V$=1.05)  \shortcite{78GRE}.
Towards  hotter main  sequence  stars, this   calibration  is valid  up to
$B_2-V_1$=0.4.   The metallicity is  related  to the index $\delta_{1256}$,
defined  as the difference between  the $U-B$ Geneva colour index
of the star and  the Hyades sequence  at the $B_2-V_1$ colour of the
star (see Grenon (1978)). This calibration is :\\

\noindent [Fe/H]$_{Gen}$=2.96+2.04/($\delta_{1256}$-0.72). \\ 

\begin{figure}
\centering                  
\begin{center}
\epsfig{file=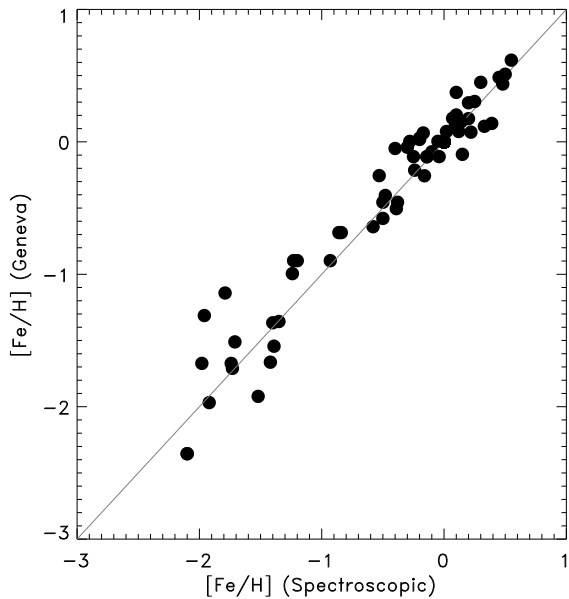,width=8.5cm,height=8.5cm}
\end{center}  
\caption{\em Photometric  determination
  of metallicity using the  calibration   by \protect\shortcite{78GRE},    versus
   spectroscopic measurements  for dwarfs with 0.40$<$ B$_2$-V$_1$ $<$  0.65.
}
\label{fig:cal-feh-gen}
\end{figure}

\begin{figure}
\centering                  
\begin{center}
\epsfig{file=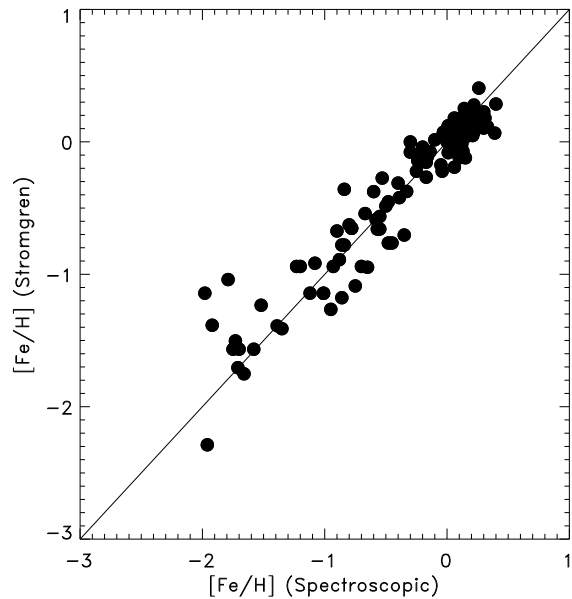,width=8.5cm,height=8.5cm}
\end{center}  
\caption{\em Str\"omgren metallicities versus spectroscopic metallicities.
Photometric metallicities are derived from the calibration by \protect\shortcite{89SCH_EA1}.}
\label{fig:cal-feh-strom}
\end{figure}   

Numerous spectroscopic metallicities have  been published for G and  K
dwarfs since  1978, so that we can  check how this calibration behaves
compared  with   recent  spectroscopic  metallicity measurements.   We
gathered  a sample  of spectroscopic metallicities  from the literature,
selecting      stars    that     have      Geneva    photometry    and
0.40$<$B$_2$-V$_1<0.65$    from      \cite{93EDV_EA2,98FEL_EA,97FLY_EA,95AXE_EA,97CAS_EA1}. We added  a sample
of Hyades  stars, with individual  spectroscopic [Fe/H], as  listed in
Perryman et al. (1997). The total sample  amounts to 76 stars.

Fig.~\ref{fig:cal-feh-gen}   shows   the    correlation   between  the
photometric  and   spectroscopic    iron abundance  in    this sample.
Dispersion increases   for   low   metallicity  stars,   however   the
photometric  metallicity  still nicely  correlates  with spectroscopic
metallicities, with no apparent  large deviations.  A least square fit
to the points of  Fig.~\ref{fig:cal-feh-gen} yields a regression curve
of  [Fe/H]$_{Geneva}$=1.002*[Fe/H]$_{Spectro}$+0.054, indicating  that
the photometric iron abundance may overestimate the spectroscopic iron
abundance by an amount of 0.05~dex.  \\

\subsubsection{Str\"omgren metallicity scale}\label{stromgrenscale}

Most previous studies  of  the  G-dwarf  problem  are  based on
Str\"omgren photometry, it is   therefore  interesting to see    how
Str\"omgren     metallicity  calibration compares       with  respect to the Geneva
photometric metallicity  and to the spectroscopic  scale.
We apply the calibration given by Schuster \& Nissen (1989) :\\

\begin{eqnarray}
[Fe/H]=1.052-73.21m_1+280.9(b-y)m_1+\nonumber\\
333.95(b-y)m_1^2-595.5(b-y)^2m_1+(5.486-\nonumber\\
41.62m_1-7.963(b-y)(log(m_1-c_3))
\end{eqnarray}

when 0.22$<b-y<$0.375	\\

\begin{eqnarray}
[Fe/H]=-2.0965+22.45m_1-53.8m_1^2-62.04m_1b-y+ \nonumber\\
145.5m_1^2(b-y)+(85.1m_1-13.8c_1-137.2m_1^2)*c_1
\end{eqnarray}

when 0.375$<b-y<$0.59.	\\

The indix c3 is defined as $c3=0.6322-3.58(b-y)+5.20(b-y)^2$\\

Fig.~\ref{fig:cal-feh-strom}
shows the spectroscopic  vs Str\"omgren photometric determination of
[Fe/H]  for  an enlarged  sample  of 110 stars,  with the Str\"omgren
metallicity    calculated  using  the  Schuster \& Nissen (1989)
calibration.  A fit applied to this sample yields 
[Fe/H]$_{Stromgren}$=0.865[Fe/H]$_{Spectro}$-0.052,   indicating a
problem in the calibration, a result similar to Alonso et al. (1996).
The correction to  this calibration  suggested by these authors is
 0.85~[Fe/H]-0.04, which  is concordant with our fit.

\begin{figure}
\centering                  
\begin{center}
\epsfig{file=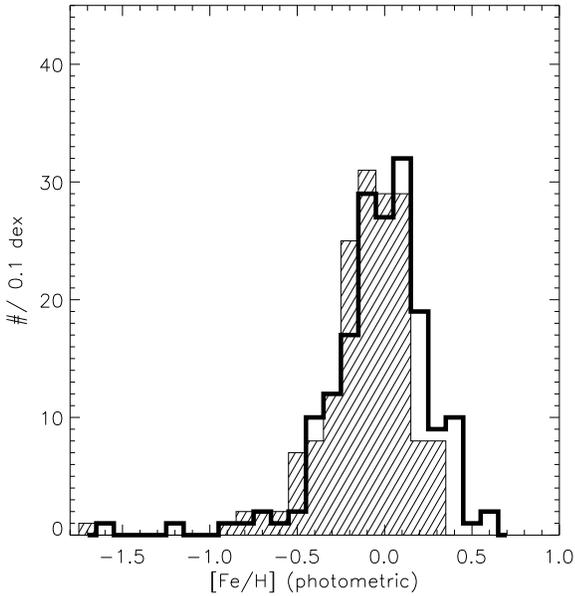,width=8.5cm,height=8.5cm}
\end{center}  
\caption{\em Stars with Str\"omgren photometric iron abundance estimates (hashed histogram) and 
Geneva photometric abundance estimates (thick line histogram).
} 
\label{histGenStro}
\end{figure}

Of the 393 stars of our sample, 324 have  Geneva photometry, 69 stars have not.  
Out of these 324 stars, 177 have 0.4 $<~$B$_2$-V$_1$~$<$ 0.65, and their metallicity can be estimated with the calibration of \shortcite{78GRE}.
Among the  147 stars with B$_2$-V$_1$ out of this interval, 
9 stars have B$_2$-V$_1$ $>$ 0.65, and are not considered any further.
We are left with 138 stars plus 69 stars which have no Geneva photometry.
Out of these 207 objects, 166 have a Str\"omgren photometry, and the corrected 
calibration above can be applied.
There are 41 stars for which no Str\"omgren nor Geneva photometry is available, and 7 of these
have a metallicity in the catalogue of Cayrel de Strobel et al. (1997). Two stars with a position
above the main sequence that suggest binarity have been removed.
Our final sample has 348 stars with estimated iron abundance.

Fig.~\ref{histGenStro} shows  the  photometric [Fe/H] distributions
of the two subsamples (Geneva and Str\"omgren photometric metallicities),  and illustrates that they cover  approximately
the same [Fe/H] range,  with a tendency for the  'Geneva' sample to be
more metal-rich. The colour criteria for the selection of the stars in
each   sample explains this difference,  the  objects  in the 'Geneva'
sample are redder, and  it is expected to favor metal-rich
stars.

\subsection{Age and mass biases}

One of the  predicted quantities of  chemical evolution models  of the
galactic disc is   the  fraction of stellar  mass   as a  function  of
metallicity. Ideally, a distribution also function of the age would be
desirable, but since measurements of the age of (most) long-lived stars is impractical,
such a distribution is still out of reach.
However, because we observe metallicity to be a rough function of age, 
an unbiased sample must not favor any particular age, and in particular 
should not include short lived stars, which are only representative of 
recent chemical evolution. 

A second bias, that has been neglected in previous determinations of the metallicity distribution, 
is introduced in observed samples by the fact that, due to selection 
criteria (e.g, colour limits or spectral types), 
the metallicity is not sampled over the same width of mass interval. 
The selection of long-lived stars and the colour 
limit imply that metal-rich stars are selected on a smaller mass interval 
than metal-poor stars. 
This would be of no consequence if chemical evolution models take this effect into account,
but this is never done in practice.
We review these 2 biases in turn.

\subsubsection{Correcting for age bias}

The selection of long-lived dwarfs is necessary to avoid bias favoring 
young stars in the sample, or recent galactic evolution.
We are looking for those stars in the sample which would still be below
the turn-off after a duration equal to the disc age.
If the (thick and thin) disc age is 12 Gyr,
this imply that the stars we are interested in have a main sequence
mass lower than the turn-off mass of the 12 Gyr isochrone at the star metallicity, which we note
M$_{TO}^*$.
In principle, because the stars in the sample may have any age between 0 and 12 Gyr, they
must be located to the right of the evolutionary track with mass  M$_{TO}^*$,
and to the left of the 12 Gyr isochrone. 

In practice, these requirements are impossible to apply because (1) we don't know the age of
the disc to a satisfactory accuracy, (2) although excellent, $Hipparcos$ parallaxes are not sufficiently accurate
to locate meaningfully the observed star between the isochrone and evolutionary track
(3) theory of stellar evolution and the various transformations necessary to compare 
the observed star with isochrones and evolutionary tracks are still too imprecise.
As a consequence, we have applied a simpler procedure which consisted in selecting
stars to the right or below the evolutionary track with mass M$_{TO}^*$. The turn-off
mass is obtained from a 12~Gyr isochrone, which is taken here as the age of the galactic (thin and thick) disc.
Lebreton (2000)   has   demonstrated   that  standard stellar  evolution
calculation shows   differences of about 100--200   K compared with the
best available  data    for deficient  and   mildly  deficient  stars.
According to Lebreton (2000), these differences can  be eliminated if (1)
non-ETL corrections  are   applied  to the metallicity   data  and (2)
microscopic  diffusion of heavy  elements  is accounted for in stellar models,
which results in effective temperature shift of about 100-200 K. 
Before selecting the stars, we artificially applied an equivalent shift 
in metallicity to the observed sample, and in temperature to 
the stellar models. 

\begin{figure*}
\centering                  
\begin{center}
\epsfig{file=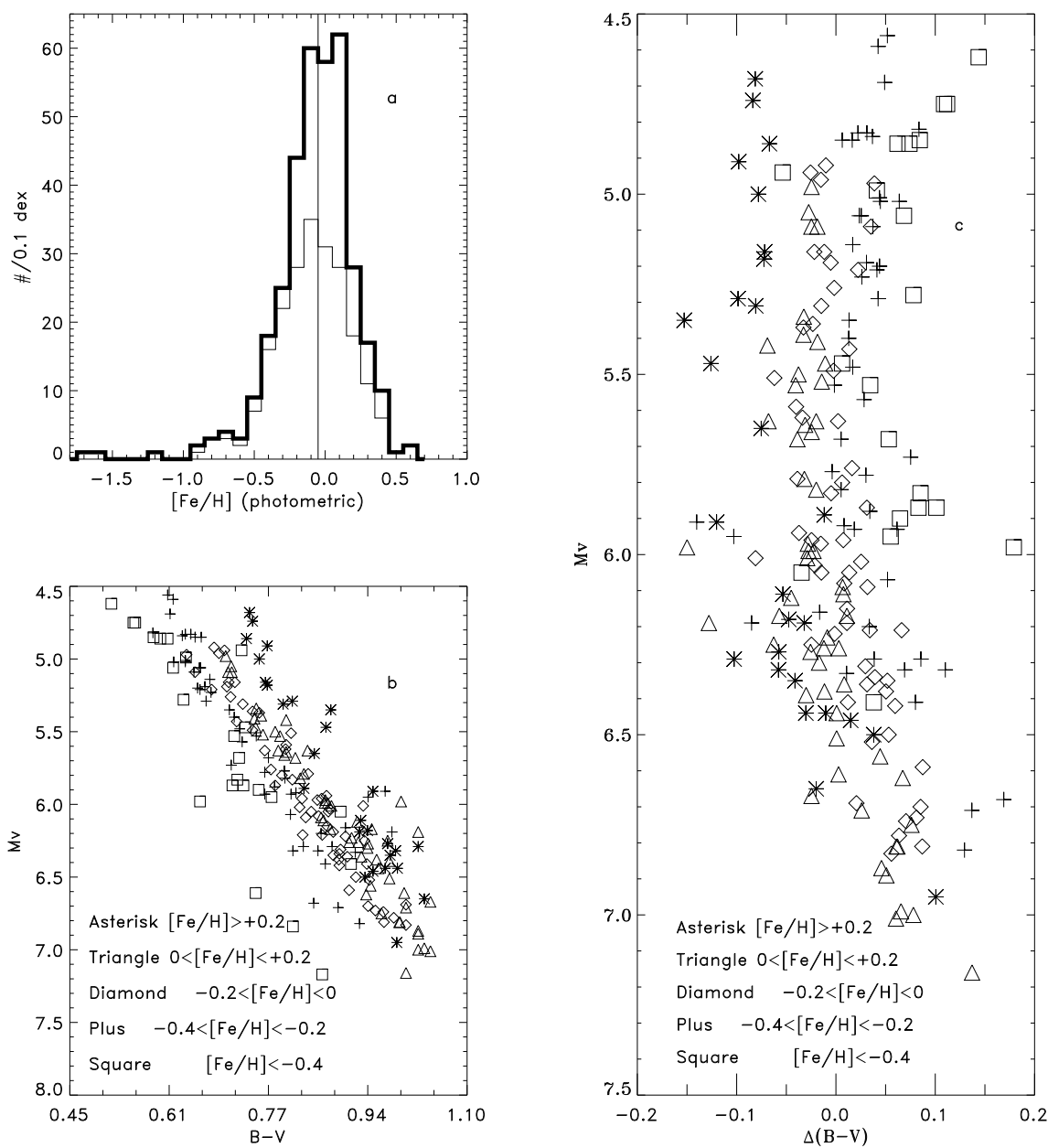}
\end{center}  
\caption{\em Description of the sample of long-lived dwarfs (a) 
Iron abundance distribution. The upper histogram (thick line) is the sample before the selection of long-lived dwarfs. 
The (thin line) histogram counts the stars left after having selected long-lived stars.
(b) HR diagram of selected stars.
(c) For a given star in the sample, the abscissa gives the $B-V$ of the solar isochrone at the magnitude of the star, minus the $B-V$ colour of the star.
}
\label{selected}
\end{figure*}

The selection of long-lived dwarfs,  applied to the 348 initial stars, yielded 218 objects.

Fig.~\ref{selected}a shows the iron abundance distribution resulting from this selection.
The initial set of 318 stars is centred on the solar metallicity, and the distribution 
of long-lived stars (218 objects) is almost symetrically centred on [Fe/H]=-0.05--0.0, a slight and expected shift.
Although the selection preferentially removes metal-rich stars, note that the effect is relatively minor.
The metallicity histogram of long-lived dwarfs has 38 per cent objects with  [Fe/H]$>$0.0, 43 per cent for the
initial sample.
The Fig.~\ref{selected}b
illustrates the position of the stars in the Hertzsprung-Russel (HR) diagram, and Fig.~\ref{selected}c shows
more distinctly the variation of metallicity along the main sequence width.

\subsubsection{Approximate correction of the mass bias}\label{massbias}

\begin{figure}
\centering                  
\begin{center}
\epsfig{file=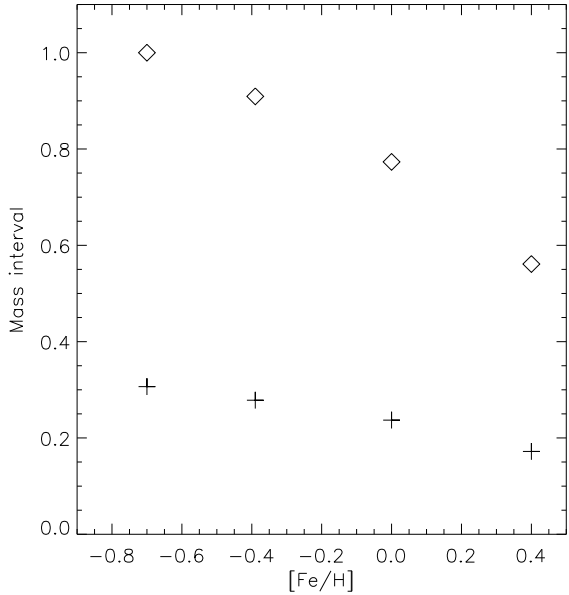,width=8.5cm,height=8.5cm}
\end{center}  
\caption{\em  Mass interval between our two limits defining our sample
on the main sequence (Turn-off and $B-V$=1.05), for different metallicities. Crosses give
the interval in solar masses. The diamonds give the fraction of mass interval relative
to the mass interval at [Fe/H]=-0.7. Correcting factor has been applied after this.
} 
\label{fig:Masscor.ps}
\end{figure}

As such, the resulting metallicity distribution of Fig.~\ref{selected}(a) should not be compared with chemical evolution 
models, because the colour limits of the sample (at $B-V$=1.05 and turn-off) introduce a bias in the
mass interval over which stars of different metallicity are sampled.
Chemical evolution models work with stellar mass percentage at a given metallicity, 
without any reference to how this mass is displayed over 'real' stars.
However, at $B-V$=1.05, stars of metallicity [Fe/H]=-0.70  have masses  of 0.60 M$_{\odot}$, whereas at  the
same   colour,  solar  metallicity  stars have  0.63 M$_{\odot}$,  and  at [Fe/H]=+0.4 
stars have masses of 0.79 M$_{\odot}$. 
 
Fig.~\ref{fig:Masscor.ps} shows how the width of the mass interval (Mass(Turn-off)-Mass($B-V$=1.05)) 
varies as  a function  of  metallicity (crosses) in stellar isochrones by \shortcite{94BER_EA}.
As   can be seen, stars   with metallicity [Fe/H]=-0.70 are
sampled over a mass  interval which is  roughly two times  larger than
those  with  [Fe/H]=+0.4.  A correction  can be  applied to the number
distribution   of  Fig.~\ref{selected}a,   which  is of  course
dependent on the amount of  stars created at a given  mass for a given
metallicity.  Our correction is a linear interpolation to the fraction
of the mass  interval  at different metallicity  relative  to the mass
interval     at     [Fe/H]=-0.7. This correction is plotted       on
Fig.~\ref{fig:Masscor.ps} (diamonds).

\begin{figure}
\centering                  
\begin{center}
\epsfig{file=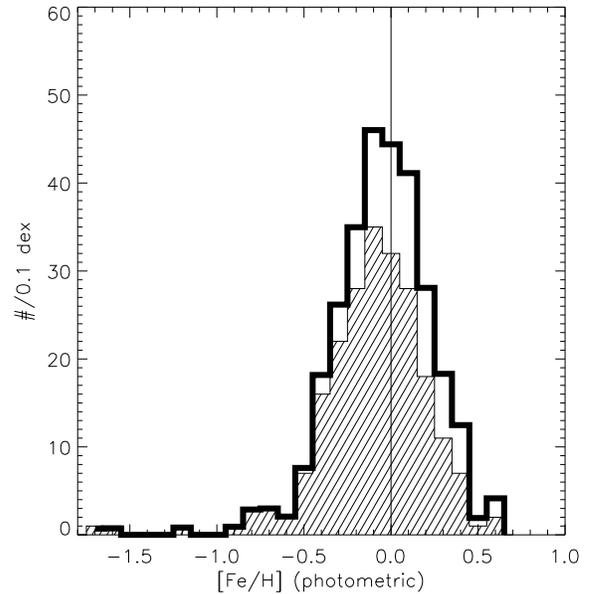}
\end{center}  
\caption{\em  Kinematically  unweighted  metallicity
   distributions of  long-lived dwarfs. The hatched  histogram is our 
   brute metallicity distribution after selection of 218 long lived dwarfs.
The thick line histogram is the distribution corrected for the mass bias, as described in the text.
} 
\label{HistFeHcomp.ps}
\end{figure}

\subsection{The metallicity distribution of long-lived dwarfs}

The effect of the mass correction on the metallicity distribution
of long-lived dwarfs is shown on Fig.~\ref{HistFeHcomp.ps}.
While strictly speaking the resulting distribution is still not comparable to chemical evolution 
models (it is not a distribution of mass fraction), it is an unbiased distribution of star numbers 
as a function of metallicity. We postpone the derivation of our final metallicity distribution of 
stellar mass to section \ref{finaldis}. 

\begin{figure}
\centering                  
\begin{center}
\epsfig{file=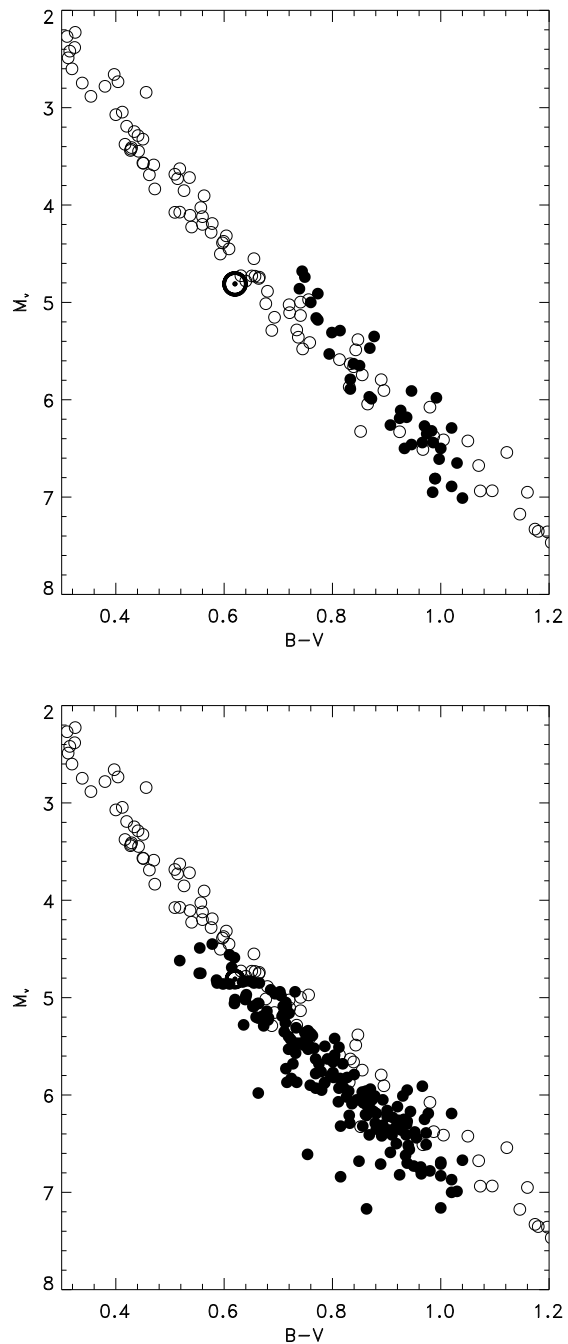}
\end{center}  
\caption{\em The sample of long-lived stars separated into 2 parts.
Plot (a) shows the Hyades sequence (open symbols) and the 40 stars in the sample that have [Fe/H]$>$+0.14, 
which is the metallicity of the Hyades \protect\shortcite{98PER_EA}, plot (b) shows stars with [Fe/H]$<$+0.14.
The Hyades stars have been selected by \protect\shortcite{98PER_EA} as single stars within 10~pc of the 
cluster centre. The sun is located at (0.62, 4.81). These figures clearly illustrate 
the existence of a large proportion of metal-rich stars in the sample.
} 
\label{stypebias}
\end{figure}

The main characteristic of our distribution is that it is centred on solar metallicity,
while previous studies have found distributions centred between [Fe/H]=-0.3 and -0.1.
In order to illustrate that our result is not an effect of a flawed  metallicity 
calibration,  we plot on Fig. \ref{stypebias} the HR diagram of our stars divided into objects
with metallicity [Fe/H]$<$+0.14 and [Fe/H]$>$+0.14, which limit is the metallicity
of the Hyades according to \shortcite{98PER_EA}. Single stars  within 10~pc of the 
Hyades cluster centre as selected by \shortcite{98PER_EA} are also plotted.
The figure clearly shows that according to their position in the HR diagram these stars
are undoubtedly metal-rich stars. 

There are two other reasons that suggest we do not overestimate the global shift of the
distribution to higher metallicities. The first one comes from a study by \shortcite{99THE_EA},
which shows that due to non-LTE effects, the spectroscopic abundance scale should be
corrected. This correction is unimportant for solar metallicities, but is of the order of
0.05~dex at [Fe/H]=-0.5.
The    second    reason   is   studied    by    \shortcite{91GIM_EA},
\shortcite{96MOR_EA},  \shortcite{97FAV_EA3}, \shortcite{98ROC_EA},
showing that chromospheric  activity affects  the photometric indice
m$_1$ on the Str\"omgren scale,  in the sense  that young active stars
may appear metal-deficient. We discuss this point in section \ref{active}.

Since previous   studies on  the  G dwarf   metallicity  have  found a
relative  consensus that  the distribution peaks  between
-0.3$<$[Fe/H]$<$-0.15, we think it is  highly desirable to spend  some
time looking  for  the origin  of the  differences  in the metallicity
distributions.

\section{Comparison with previous metallicity distributions}\label{section_bias}

\subsection{General comments}

Different biases   may   affect the metallicity    distribution due to
selection criteria of  the observed stars. Two  such  biases have been
discussed and corrected  in the previous  section.  In the literature,
only the first one  has been corrected  (the age bias), although it is
the least  important one.  An even more  problematic bias occurs prior
to the two  former biases mentioned, in samples  selected on the basis
of spectral type. A selection on spectral type is sometimes imposed by
the availability of photometric  surveys (such as  Str\"omgren surveys
of Olsen), but  seems to be perpetuated  for rather historical reasons
(the `G-dwarf' metallicity  distribution) than real limitations in the
available    data.  Unfortunately, spectral   type criteria introduce
biases in the metallicity  distribution which are almost impossible to
correct.

We have demonstrated in section \ref{massbias} how the colour limit of
the  observed   samples   introduces  a non-negligible    bias against
metal-rich stars.  In order to be able to correct  this bias, one must
know the colour limits  of the sample.   This is impossible to  obtain
once the selection has been made from  spectral type, due to the intrinsic
high  dispersion  of  colours at   a  given  spectral type.    A brief
inspection  of  the Third  edition of the   Catalogue  of Nearby Stars
(hereafter  CNS3, \shortcite{91GLI_EA})  illustrates this
point: for  example G5 V stars  have colours extending from $B-V$=0.58
to 1.1.  and G0 V stars from 0.50 to 0.69. The intricacy of colours
between spectral types forbids a  clean definition of the sample  mass
limits. 

The  metallicity bias introduced is even  more severe if  one does not
include sufficiently   late   spectral types.  This  point    has been
discussed   in   \shortcite{78GRE}, \shortcite{87GRE}.  According to
\shortcite{90GRE2},    turn-off    of   the oldest solar metallicity  stars is
$B-V$=0.68,  which  implies  that samples selected   on spectral types
earlier than G2 V  or G5 V  would miss most of them,  not to speak  of
metal-rich stars. We show below that  an unbiased sample must comprise
spectral type as late as K. 

In the  last  five years numerous studies   of  the local  metallicity
distribution    of          dwarfs         have   been       published
\shortcite{95WYS_EA,96ROC_EA3,97FLY_EA,97FAV_EA2,98ROC_EA}.   All are based  on
trigonometric distances from the CNS3 or CNS2. The first two and \shortcite{98ROC_EA} used
Str\"omgren photometry, while   Flynn  \&  Morel  designed their   own
metallicity indicator based on the $B_1$ Geneva  and the $R-I$ Cousins
colour  indices.  Finally \shortcite{97FAV_EA2}  relied on  their own
spectroscopic determinations.  As a general rule,  it can be said that
the  metallicity distributions of all these   authors present the same
characteristics.     Briefly  :  a  broad  maximum    is found between
[Fe/H]=$-0.3$   and   $-0.1$,   20--30 percent  of    the   stars have
[Fe/H]$>$0.0, 10 to 20 per cent between -1.0$<$[Fe/H]$<$-0.5. 
Very few or no  stars have [Fe/H]$<$-1  (however Flynn \& Morel find 7
stars with [Fe/H]$<$-1.0) or [Fe/H]$>$0.3.  The exception to this rule
is \shortcite{97FAV_EA2}, with    a  distribution that peaks    at  a
somewhat higher metallicity than  the others with the consequence that
41 per cent of their stars have a metallicity higher than solar, and 7
per cent between -1.0$<$[Fe/H]$<$-0.5. 

In order to understand the origin of  the differences and similarities
between  our  metallicity   distribution    and those   obtained    by
\shortcite{95WYS_EA,96ROC_EA3,97FAV_EA2},  we examine in more   detail
the samples used by these authors.

\subsection{Comparison with other studies}

\subsubsection*{\protect\shortcite{95WYS_EA}}

Out of 90 stars in  the Wyse \& Gilmore (1995)  sample  given as nearer  than
30~pc in the  CNS3  catalogue, 28 are  in  fact at a   larger distance
according to $Hipparcos$ parallaxes.   The great majority of  these were
selected   on the  basis  of   trigonometric   parallax in the    CNS3
corresponding to a distance between 20 and 30~pc. All had large errors
on parallaxes%
\footnote{They provide a clear illustration of a
Lutz-Kelker effect on trigonometric parallax. We note in passing that 18  of these stars 
also have a  spectroscopic or photometric  parallax in the CNS3,  which  in most   cases is  much nearer  to the
$Hipparcos$ parallax  (to   within 10 per cent for   most).}.

An interesting point is the presence of metal-rich stars in our sample
which are  totally absent from  the sample  by  Wyse \& Gilmore.  This
difference is an illustration of the bias introduced by spectral type.
The     sample   of   Wyse \& Gilmore (1995)  resulted    from   the
cross-identification of  the CNS3 and \shortcite{OLS}  catalogue of
Str\"omgren photometry for  F/G0 spectral types. Wyse \& Gilmore (1995)
then applied   a 2$^{nd}$ selection  by   excluding stars outside  the
interval 0.4$<B-V<$0.9.   A consequence of their 
spectral type
selection is that 10 stars in our
sample with  [Fe/H]$>$ 0.2  and 
0.4$<$ B-V $<$0.9 
are  absent from the Wyse \& Gilmore (1995)
 sample.  Nine of these objects  are in the CNS3,
but have  G5V--K0V spectral types, and  are therefore absent  from the
catalogue  of \shortcite{OLS}.   Even more  surprising  is the fact
that for 39  stars with [Fe/H]$>$0 and $B-V<$0.9  in our sample,  none
has  a (Simbad) spectral  type earlier  than  G5.  Therefore, 39 stars
selected as long-lived objects and [Fe/H]$>$0 in our sample are absent
from the sample of Wyse \& Gilmore (1995). 

Finally, note that   due to the  accurate  $Hipparcos$ parallaxes, the
overdensity  of  metal-poor stars   found by Wyse \& Gilmore (1995)  is
washed out.  Fifteen stars ($\approx$  6 per cent) have $[$M/H$]<$-0.5
in our sample, 17 (19  per cent) in Wyse \& Gilmore (1995).  Six  stars
only are in common. Nine stars in  the sample of Wyse \& Gilmore (1995)
have a parallax smaller than 40 mas and  have not been included in our
sample. Two stars have   a parallax larger than   40  mas but have   a
multiplicity flag in the $Hipparcos$ catalogue.

\subsubsection*{\protect\shortcite{96ROC_EA3}}

\begin{figure}
\centering                  
\begin{center}
\epsfig{file=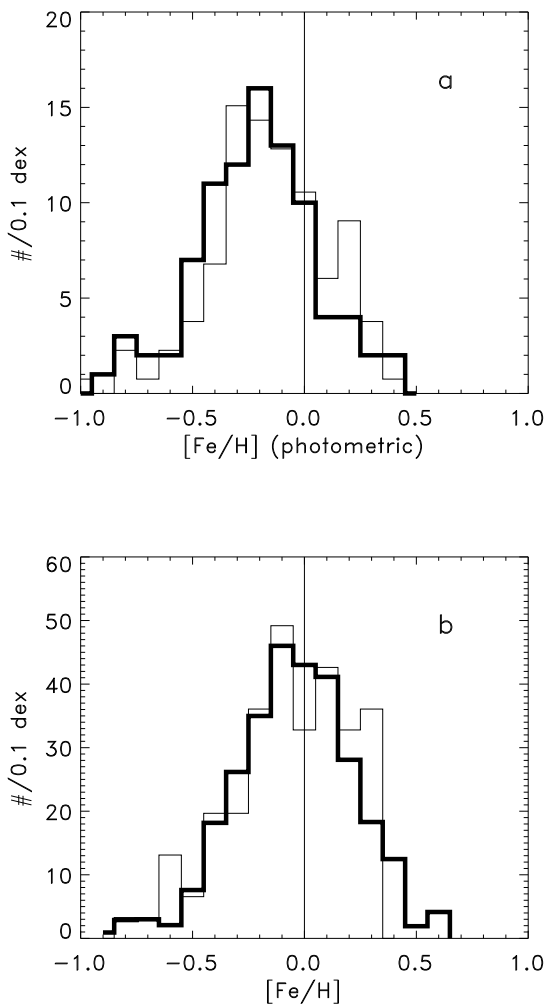}
\end{center}  
\caption{\em Plot (a) illustrates  how a bias against metal-rich stars
is introduced by selecting G-type stars.  The plot shows our sample of
long-lived dwarfs,  with   all K-type stars   removed,   to mimic  the
metallicity     bias   (in        thick   line).    The    sample   of
\protect\shortcite{96ROC_EA3},  which  contains  only G-type    stars
(cleaned using $Hipparcos$ parallaxes) is also  shown.  The lower plot
shows  our distribution (all stars,  corrected as described in section
2) and the distribution by \protect\shortcite{97FAV_EA2}, which shows
no apparent bias.  Histograms have been normalised to contain the same
number of stars.} 
\label{sampleST}
\end{figure}

\shortcite{96ROC_EA3} have  selected  a sample   similar  to that  of
Wyse \& Gilmore (1995), except that  they  extended their selection  to
all G dwarfs in  the Gliese catalogue.  Thus,  they are less sensitive
to bias against metal-rich stars (but see below): approximately 20 per
cent of their  sample has [Fe/H]$>$0.   They find very  few stars with
[Fe/H]$>$0.2 (approximately 2  per cent), and their distribution peaks
between  [Fe/H]=-0.2 and -0.3.    Several  factors contribute to   the
importance of this peak. First note that 47 per  cent of the stars are
illegitimate   in the  sample of    \shortcite{96ROC_EA3}, because of
parallaxes smaller than  40~mas according to $Hipparcos$.  Second, the
use   of \shortcite{89SCH_EA1} metallicity calibration underestimates
the metallicity of  the stars.   Finally,  the limitation to  spectral
types  G  is undoubtedly  a  source of  bias,  (even though it is less
important than  for the previous study).  
Fig.~\ref{sampleST}(a)  shows their   metallicity distribution,  after
cleaning from objects with $\pi_{Hipparcos}<40$~mas. We also removed those objects
which are subgiants and  giants according to  their position in the HR
diagram.  The   result  is  not   very  different  from  their initial
distribution (see their Fig. 2). 

On the same plot, we show  our sample, from  which we have removed all
K-type     stars,   in   order    to       mimic  the  selection    of
\shortcite{96ROC_EA3}.  The effect of  this simple modification is to
produce a new distribution   which peaks at  -0.3$<$[Fe/H]$<$-0.2,  in
agreement with that of \shortcite{96ROC_EA3}. The  bias is clear : by
simply  removing   K-type  stars,   the   sample    is  shifted   from
[Fe/H]$\approx$0.0 to -0.3$<$[Fe/H]$<$-0.2

\subsubsection*{\protect\shortcite{97FAV_EA2}}

The distribution obtained by Favata et  al. is interesting because (1)
it is  based   on spectroscopic  measurements and  (2)  its maximum is
shifted towards solar metallicity stars, as in  our sample.  They have
measured the iron abundance for 91 stars among the 1979 edition of the
catalogue of Nearby stars (Gliese \& Jahreiss, 1979), limited to stars
with $\pi>$ 0.045$\arcsec$.    The distance definition   of the sample
suffers from problems mentioned for previous samples, with 40 per cent
of these 91 stars being outside the 22.5~pc sphere.  However, although
the sample is  obviously incomplete in  distance,  it is probably  not
biased by  spectral type because the objects  were selected at random
in a given colour interval.  The result on Fig. \ref{sampleST}(b) is a
distribution very similar to that of our sample.

\subsubsection*{Conclusions}

Examination of the first two  examples shows that they certainly suffer
from biases introduced by a selection  on spectral types. Although the
attention  in    the literature has   been mainly    focused on biases
affecting   low metallicity  stars, we argue    that at  least as much
important biases have truncated   the metal-rich ([Fe/H]$>$0)  side of
the  metallicity   distribution.  All  three  samples  are  plagued by
incompleteness of the colour intervals and underestimated distances from
the Catalogue of Nearby Stars. 
As a conclusion, we should say that the two studies by \shortcite{96ROC_EA3} and
Wyse \& Gilmore (1995) have provided G dwarf metallicity distributions consistent with each other.
However, G dwarfs are not representative of the metallicity of the disc,
and are biased against metal-rich stars. On the contrary, when the samples are 
not limited to G spectral types, as is the case for \shortcite{97FAV_EA2},
the metallicity distribution contains about 40 per cent of stars with [Fe/H]$>$0.0.

\subsection{A glimpse on the age distribution of the sample}\label{active}

In models describing the chemical  evolution of the galactic disc with infall, the
infall rate fixes the pace   at which the   galactic disc is  built,
through the   SFR.  This is well  illustrated  in  models  reviewed by
Tosi (1996).   Because the  infall  rate is usually  taken as a
decreasing  function of time,  the SFR   history follows. Thus,  since
infall  is so widely  used  to explain the   lack of metal-poor dwarfs
observed in the solar neighbourhood, a  desirable test of these models
would  be the  age distribution  of dwarfs  that make the  metallicity
distribution.   Unfortunately, as mentioned   previously, most of  the
stars in the sample have  a main sequence evolution  that is too close
to the ZAMS to  have an age  determined with some  accuracy in  the HR
diagram. However, an  interesting   information is available  for  the
young dwarfs because they can be detected as X-ray emitters
due to coronal activity.  The cross-identification between the $ROSAT$
survey and the Catalogue of  Nearby Stars \cite{99HUE_EA} can be  used to quantify the
percentage of potentially young stars in our sample.

\begin{figure}
\centering                  
\begin{center}
\epsfig{file=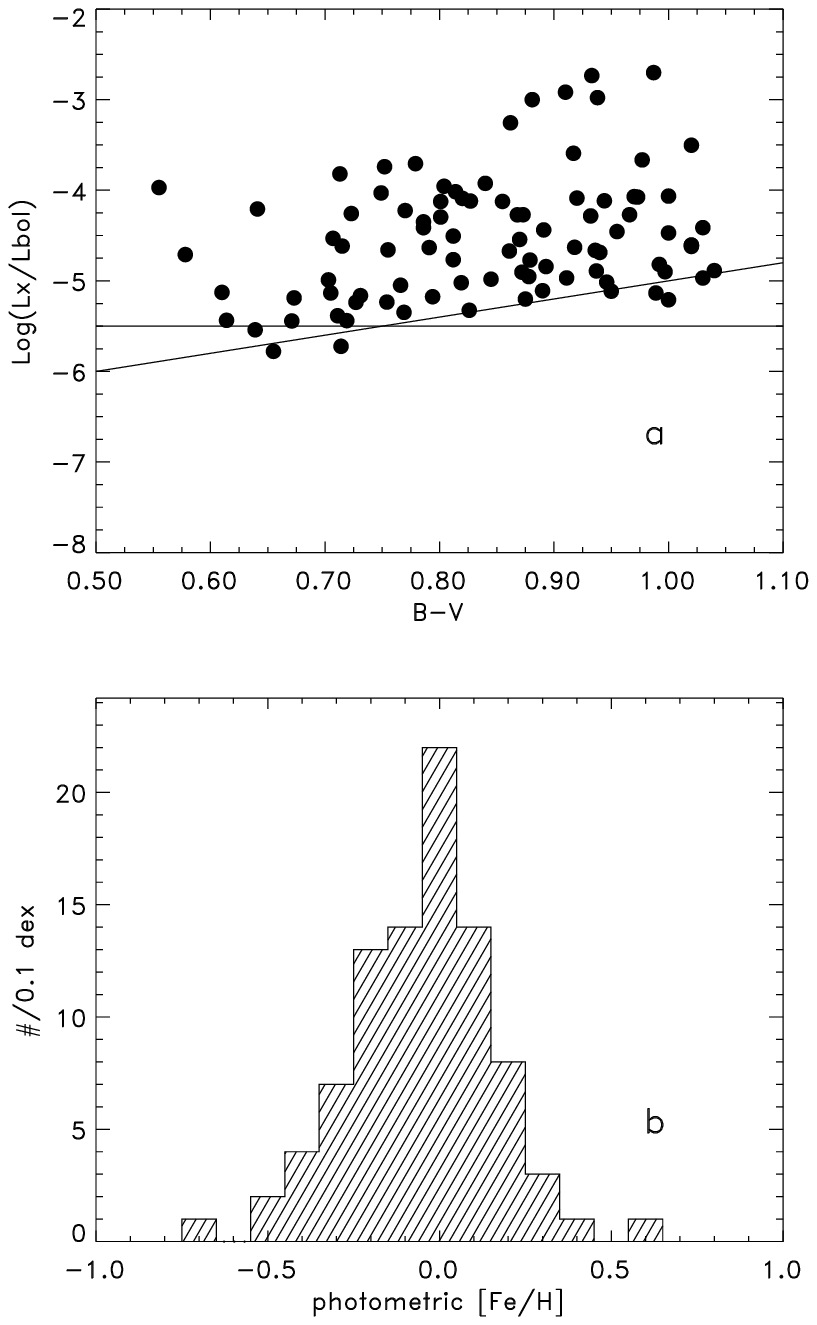}
\end{center}  
\caption{\em 
(a) X-ray flux as given in \protect\shortcite{99HUE_EA} as a function of $B-V$ colour.
The horizontal line shows approximate flux level for 2 Gyr stars according to the logL$_X$-logR'$_{HK}$
relation of \protect\shortcite{97STE_EA} and the logR'$_{HK}$-logt relation of \protect\shortcite{91SOD_EA}.
The other line delineates the incompleteness of the X-ray data, according to \protect\shortcite{97STE_EA}. 
The 90 stars may slightly underevaluate the number of X-ray emitters in our sample, 
because of incompleteness of the X-ray data $B-V>$0.8.
(b) Photometric metallicity distribution of stars in (a).}
\label{nainesxbv.ps}
\end{figure}

There are 90 stars (40 per cent) in our sample in the list of objects resulting from this cross-identification.
Fig.~\ref{nainesxbv.ps}(a) shows the X-ray   flux of these stars  as a
function of $B-V$ colour. The figure shows that our sample may contain slightly
over   90   stars      with   X-ray     luminosity  brighter      than
log~X/L$_{bol}$=-5.5, which could be absent from the sample of Huensch et al. (1999),
due to the incompleteness to the X-ray data for the reddest stars.
In order to get a rough estimate the corresponding ages, we utilise the log~L$_X$/L$_{bol}$-logR'$_{HK}$
relation of Sterzik \& Schmitt (1997) and the logR'$_{HK}$-logt relation of Soderblom et al. (1991).
According to these relations,  log~L$_X$/L$_{bol}$=-5.5 corresponds to an age of
2 Gyr. Most our stars having X-ray emission at the level of log~L$_X$/L$_{bol}>$-5.5  could
therefore be considered younger than 2 Gyr.

Fig.~\ref{nainesxbv.ps}(b)  illustrates  what    the metallicity
distribution of these stars is. It is centred on [Fe/H]=0.0, with an
unexpected but important contribution of  11 stars with [Fe/H]$<$-0.3.
Some of these stars have a spectroscopic iron abundance which confirms
the    photometric iron abundance.     For   instance, HIP 88622   has
[Fe/H]$_{photo}$=-0.506, and [Fe/H]$_{spectro}$=-0.47, dated 15~Gyr in
\shortcite{93EDV_EA2}, and  HIP 15510 has [Fe/H]$_{photo}$=-0.35, and
[Fe/H]$_{spectro}$=-0.48 \cite{94PAS_EA}.   The X-ray  emission of HIP
88622 may  appear somewhat  puzzling because  of  its metallicity, and
because HIP 88622  has no significant chromospheric emission according
to Pasquini et al. (1994).  While these cases may be exceptions, it has
been suggested  more  generally that activity  may  affect photometric
indices and  that presumably young  stars may be measured as deficient
objects      if   abundance  is    measured    from   photometry 
\cite{91GIM_EA,96MOR_EA,97FAV_EA3,98ROC_EA}. This effect may explain  the shape of    the
histogram of Fig.\ref{nainesxbv.ps}, but its importance is
uncertain, and we do not  try to correct it.  While  some of the stars
may have been detected as X-ray emitters in the sample for reasons not
related to their age, it  is clearly demonstrated that X-ray  emission
is  a tracer of  young stars  \cite{98GUI_EA}.
In this  respect, the fact that  around 40 per  cent  of the sample is
composed  of X-ray emitters  that have ages less than  2 Gyr suggests a
few comments. 

In section \ref{models}, we calculate a model metallicity distribution
assuming a constant SFR.  We argue that although a constant SFR is not
phenomenologically directly   related    to the gas    content,  it is
compatible  with  available determination  of the SFR  history  in the
Milky Way.    If a  constant  star formation  rate has  dominated  the
history of the galactic disc, then of the order  of 15--25 per cent of
the stars are expected to have an age less than  2 Gyr (for a 8-12 Gyr
thin    disc).   However, a   distance  limited  sample  of  the solar
neighbourhood  is biased  against old  stars   because of the  secular
heating of the disc, observed in the increase of the vertical velocity
dispersion    with age.  Using  the   correction factors introduced in
section~\ref{scaleheight},   (factor  of 5   decrease  in the  surface
density for  stars between 8  and 4~Gyr, 2.2  for stars between  4 and
1~Gyr, normalised to  1 for stars with age  less than 1~Gyr.  The thin
disc age is taken to be 8~Gyr;  adopting 10~Gyr, this estimate changes
to 40 per  cent), we obtain of the  order of 45  per cent of the local
stellar material  to have an age  less than 2~Gyr.  This is compatible
with 40 percent of the stars having X-ray emission.

\section{Final distribution}\label{finaldis}

\subsection{ [Fe/H] distribution}

\begin{figure}
\centering                  
\begin{center}
\epsfig{file=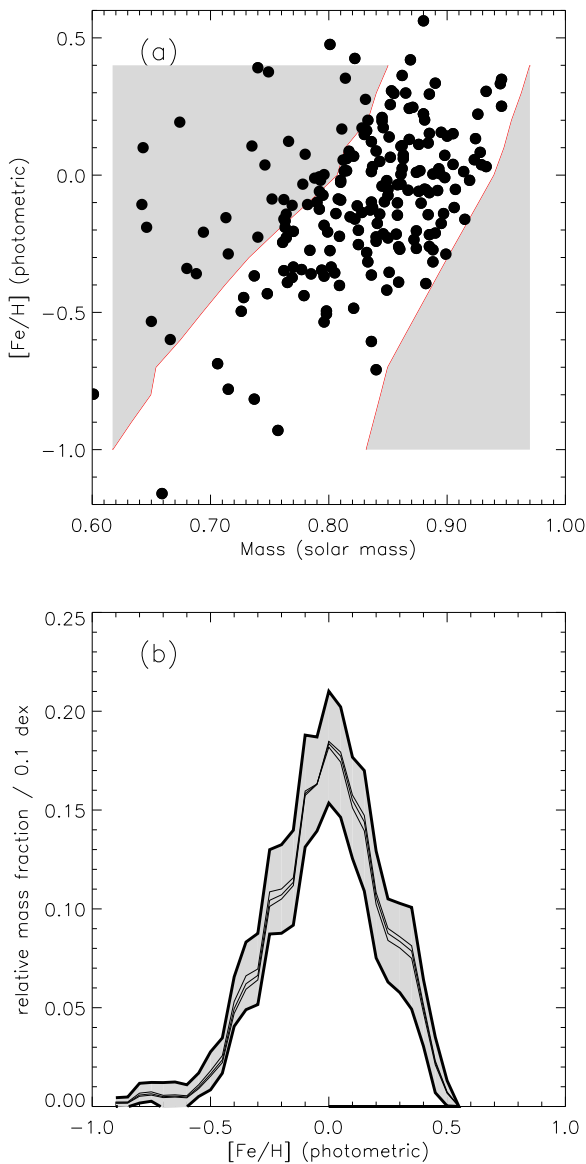,height=16cm}
\end{center}  
\caption{\em 
({\bf a}) : Mass-metallicity distribution for dwarfs in the sample. The 2 lines show the selections
due to the colour limit at $B-V$=1.05 (left) and turn-off (right). The gray areas shows the limits
within which a correction has been applied.
({\bf b}) : the final metallicity distribution of stellar material in the solar neighbourhood.
The result is illustrated adopting 3 different IMF slopes (0.05,0.7,1.2), which shows negligible
effect (the 3 curves). The gray area shows the uncertainty in the distribution due to poisson noise.
}
\label{massmetal}
\end{figure}

As already  mentioned, a  distribution  suitable for  comparison  with
presently available chemical evolution  models  is the  percentage  of
mass found at different metallicities.   To obtain such  distribution,
we estimate the  mass  of   each star  from  the  adequate   isochrone
\shortcite{94BER_EA}  --  for  unevolved  stars -- or evolutionary
track -- for those   stars for which evolution  is   significant.  The
masses of unevolved stars have been estimated by matching the observed
and theoretical absolute magnitudes   on the isochrone.  We  have  not
taken the  colour  into  account  because  of  the uncertainty of  the
stellar  model  temperatures      and  the   conversion   to    colour
\shortcite{00LEB}.  For this reason,  it is expected that for some
stars that fall  below their isochrone, the  mass may correspond to  a
colour on the  isochrone that is  beyond the colour limit at  B-V=1.05
shown on   Fig. \ref{massmetal}a.  Stars  near  the turn-off  all fall
within the limit, because the  fit to the  stellar models includes the
age as a supplementary  parameter.  The resulting fraction  of stellar
mass as a function of metallicity is plotted on Fig. \ref{massmetal}b.

\begin{table}
 \centering
  \caption{Percentage of stellar mass as a function of metallicity in the solar neighbourhood
 per 0.1 dex bin}
  \begin{tabular}{@{}lllr@{}}
[Fe/H]&   \%  &  error(\%) \\
-0.90 &  0.23 &  0.22\\
~   -0.85 & ~ 0.23 & ~ 0.25\\
-0.80 &  0.68 &  0.50\\
~   -0.75 & ~ 0.74 & ~ 0.48\\
-0.70 &  0.57 &  0.64\\
~   -0.65 & ~ 0.59 & ~ 0.65\\
-0.60 &  0.54 &  0.56\\
~   -0.55 & ~ 1.10 & ~ 0.59\\
-0.50 &  1.80 &  0.96\\
~   -0.45 & ~ 2.58 & ~ 0.89\\
-0.40 &  5.31 &  1.25\\
~   -0.35 & ~ 6.61 & ~ 1.71\\
-0.30 &  6.96 &  1.80\\
~   -0.25 & ~10.86 & ~ 2.13\\
-0.20 & 11.01 &  2.24\\
~   -0.15 & ~11.58 & ~ 2.40\\
-0.10 & 15.96 &  2.83\\
~   -0.05 & ~16.32 & ~ 2.38\\
 0.00 & 18.18 &  2.83\\
~    0.05 & ~17.43 & ~ 2.79\\
 0.10 & 15.11 &  2.57\\
~    0.15 & ~13.95 & ~ 3.05\\
 0.20 & 10.22 &  2.70\\
~    0.25 & ~ 8.40 & ~ 2.10\\
 0.30 &  8.02 &  2.26\\
~    0.35 & ~ 7.50 & ~ 2.58\\
 0.40 &  4.74 &  1.69\\
~    0.45 & ~ 2.11 & ~ 1.41\\
 0.50 &  0.67 &  0.64\\
\end{tabular}
\label{distrib}
\end{table}

In order  to correct for the bias  introduced by the colour  limit and
the selection of long-lived  objects, we use the  following procedure.
We assume a given initial mass function, the same at all metallicities
over  the mass interval   of interest, from  0.60  to 0.95M$_{\odot}$.
This IMF is used to extrapolate the amount of stellar mass outside the
mass interval where completeness is achieved.   The number of stars as
a function of [Fe/H]  within the mass  interval where  completeness is
achieved is used as  normalisation.  The plot of fig. \ref{massmetal}b
shows the resulting metallicity distribution for different IMF slopes.
The influence of the IMF  slope is negligible,  due to the small  mass
interval  involved.  We have  also  evaluated the  contribution of the
poisson noise to the distribution. If N stars in the sample contribute
to  a given metallicity  interval $\Delta$[Fe/H], we  can evaluate the
total mass $\sqrt N$ stars  would  represent, weighted by the  adopted
IMF.  The resulting distribution, together with its errors is given in
Table \ref{distrib}. 

\subsection{[O/H] distribution}

In order to have a distribution comparable with the predictions of the
Simple Closed-Box  model with instantaneous recycling approximation or
other types of models which  use the instantaneous approximation,  the
iron  distribution  is usually   converted   to  an  oxygen  abundance
distribution.  Since Pagel (1989), this  is achieved by noting
that  oxygen, being mostly  synthesized in  type II supernovae,  whose
recycling time is  negligible,  is better suited  to compare  with the
Simple  model.  Clegg et al. (1981) have shown that [O/H]=0.52[Fe/H]+0.03
for -1$\le$[Fe/H]$\le$+0.4, which  has  subsequently been approximated
to [O/H]=$0.5$[Fe/H] in  studies dealing with  the G-dwarf metallicity
distribution. Applying  this relation to  the [Fe/H] distribution much
strengthens the  narrowness  of the  distribution  and the disagreement
with the SCB  model, and it deserves  more attention.  There is  now a
large debate on the  exact behavior of the  [O/Fe] ratio as a function
of  [Fe/H]  for   [Fe/H]$<$-1.0.  We   are   interested  here in  the
metal-richer part at [Fe/H]$>$-1.0. 

When measured on the dataset by Edvardsson et al. (1993), a linear regression yields 
\begin{equation}
[O/H]=$0.64$*[Fe/H]-0.041
 \label{Fe2O1}
\end{equation}

while the study by \cite{00CAR_EA} gives the relation
\begin{equation}
[O/H]=$0.51$*[Fe/H]-0.064
 \label{Fe2O2}
\end{equation}

and the study by \shortcite{00CHE_EA} gives the relation
\begin{equation}
[O/H]=$0.70$*[Fe/H]+0.07
 \label{Fe2O3}
\end{equation}

\begin{figure}
\centering                  
\begin{center}
\epsfig{file=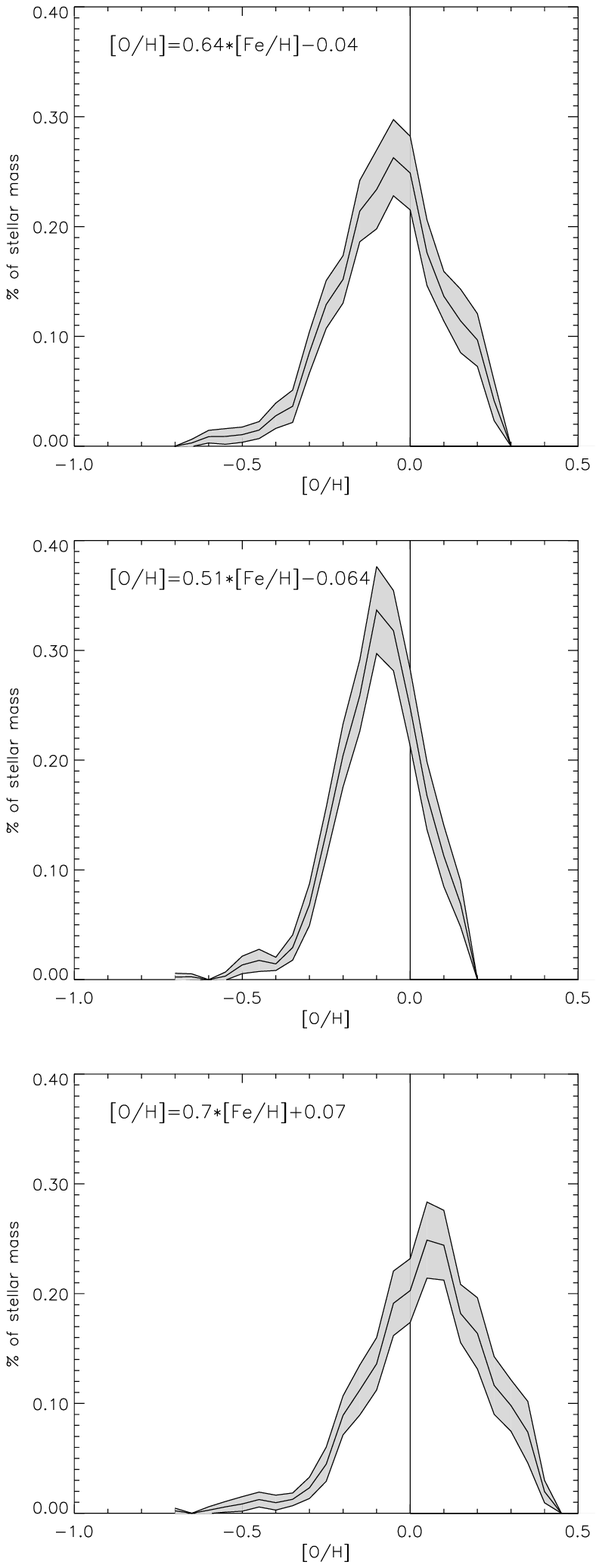}
\end{center}  
\caption{\em Distribution of stellar mass with [O/H], adopting 3 different [Fe/H] to [O/H]
relations from \protect\shortcite{93EDV_EA2} (a), \protect\shortcite{00CAR_EA} (b), \protect\shortcite{00CHE_EA} (c).
} 
\label{massoh}
\end{figure}

The difference of 0.11 dex offset between the relations \ref{Fe2O1} and
\ref{Fe2O3} is  due to different temperature scales  adopted in the two
studies,  as mentioned  by Chen et al. (2000).   In the metallicity
range of interest,  the  second relation  gives similar results  to the
relation  of Clegg et al. (1981),    implying   a  narrow    [O/H]
distribution,  but with  a  shift of 0.1~dex.   On   the contrary, the
relation of Chen et al. (2000)   seriously reduces the  effect of
converting  from [Fe/H] to  [O/H].    Looking at Fig~\ref{massoh},  it
seems that the conversion to  [O/H] comes as an additional uncertainty
in an already  long list of approximations whose  effect has  not been
seriously accounted for.  There  are too few indications  that
can help  us deciding which conversion to apply, hence, we  prefer to
keep on working with the [Fe/H] distribution.

\section{The Simple Closed-Box model}\label{models}

Two  kinds of assumptions   define the Simple   Closed-Box model.  The
first kind assumes that  complex physical processes can reasonably  be
represented  by  simple analytical    laws.  They reflect our   (poor)
knowledge of these processes but are convenient  for the derivation of
the model results.  These assumptions  include  : (1) an IMF  constant
with time, (2)  no exchange of  matter (3) the interstellar medium  is
initially free of metals (4) the interstellar medium is homogeneous at
all times. 
Assumptions of  the second kind are  introduced  purely for analytical
tractability.  This is   the   case for  the  instantaneous  recycling
approximation (IRA).   We are tempted to classify   assumption (4) in the
second category,  because it  is  possible to conceive  sub-SCB models
which would evolve independantly at slightly  different rates and that
would preserve individually the characteristics of the SCB model. This
would  give rise  to some  amount of  inhomogeneity in the metallicity
distribution without corrupting the  Simple model as the main evolutionary path
for chemical evolution.

\subsection{The `G-dwarf problem' : A little bit of semantics}

The `G-dwarf problem' is usually referred to as the lack of metal-poor
stars in the vicinity of  the sun, relative to the SCB
model.  However,  the `G-dwarf problem'  seems to cover  two different
problems.   In Section \ref{section_bias}, we  have demonstrated that there
is a real   `G-dwarf' problem in the  sense  that a sample  of G-dwarf
stars  is  inherently biased in  regard to  metallicity, and cannot be
compared directly to chemical evolution models. Perhaps the `G-dwarf problem'
should more appropriately only refer to this specific bias.

The second  aspect of  the so-called  G-dwarf problem  (which arguably
should better  be quoted  as  the SCB model   problem), refers to  the
long-lasting difference  between  the  amount  of metal-poor  material
generated by  the     SCB   model, and    the   observed   metallicity
distribution.

\subsection[]{Model input parameters}\label{parameters}

The derivation of the metallicity distribution of the stellar material in the
case of the SCB model with IRA can be made analytically, due to the 
assumptions of the model, see for example \shortcite{87BIN_EA}. In the case of
zero initial metallicity of the gas in the disc, the metallicity evolves 
as \\

\begin{equation}\label{equ_scb}
Z(t)=-p\-ln[S_{gas}(t)/S_{gas}(0)]
\end{equation}

where S(0) is the initial gas surface density, and S(t) is the gas surface density at time t.

In the above equation, the yield  p modulates the  position of the peak of
the  metallicity distribution, and  S(t) is an implicit combination of
the SFR consumption of gas and the rejection rate of the stars.  It is
usually chosen so that the final ratio S$_{gas}$(t)/S$_{gas}$(0) meets
the (loose) observational constraints of the solar neighbourhood. When
the   instantaneous approximation is not   assumed,  the gas  fraction
evolves as  the complex result  of stellar  ejecta, lifetimes, and gas
definitively locked up  in low mass stars  and  stellar  remnants.  
The  resulting metallicity  distribution is a function
of the yield and  the gas fraction  of the disc, while  the comparison
with the data  also depends on the disc  thickening. We review each of
these three points in turn.

\subsubsection[]{The yield}\label{yield}

The yield of a  generation of stars is the  mass of metals synthetized
and ejected per  unit mass of material  locked for a sufficiently long
time    in   stars and  stellar    remmants.     It is  dependant   on
nucleosynthesis, mass loss rate and  stellar ejecta,  but also on  the
IMF.  In  the case of the  SCB model with  IRA, the yield is chosen so
that the equation above is compatible with the present fraction of gas
in the disc  and abundance of the  interstellar medium.  The abundance
of   the   interstellar  medium is   usually   evaluated  to  be solar
\shortcite{98BIN_EA} (this parameter is however very uncertain, 
and  not necessarily
representative of the  evolution on the  last Gyr.   
).
The gas fraction is situated between 0.1-0.3, depending on the adopted
gas  density   (6--11M$_{\odot}$.pc$^{-2}$)  and     total     density
(40--50M$_{\odot}$.pc$^{-2}$). With  these   values and equation \ref{equ_scb},   the  yield   is
evaluated to be within the range 0.0086-0.025, which is a fairly large
range, and is of little help in defining the paramaters of the SCB model
with IRA.

How does this estimate compare  with the yield  calculated from  the local
IMF and stellar yields ? Such yields have been calculated using a model
with the  following  characteristics :  The gas  is ejected from stars
assuming the  same initial to  final   mass relation  as that used  in
Scully et al. (1997), which comes from Iben \& Tutukov (1984) :\\ 

 m$\le$6.8M$_{\odot}$ ~~M$_R$=0.11$\times$m+0.45 M$_{\odot}$ 

 m$>$6.8M$_{\odot}$ ~~M$_R$=1.5M$_{\odot}$. \\

The stellar yields  come from Portinari \& Chiosi (1999) for massive stars
and van den Hoek \& Groenewegen (1997) for intermediate mass stars.

Although  the   IMF  of solar   neighbourhood stars    is still highly
uncertain, it is   now admitted that  it  has at   least two different
regimes,  for   low (roughly lower  than one   solar  mass),  and high
masses. The IMF at low  masses  is essential,  but  in many cases in the literature the
value   of  the  index   adopted in  chemical  evolution  models seems
unrealistic. For instance,  Portinari \& Chiosi (1999) adopt  x=1.35  for
M$<$   2    M$_{\odot}$,     while others  Gratton et al. (2000) use
Scalo (1986), which is also too  steep at low masses. Other works
utilise  a single   index    for calculating   the yield,     such  as
Pilyugin \& Edmunds (1996).  Recent measurements  of the IMF from the solar
neighbourhood luminosity function have proved that the IMF index at low
masses  is    much shallower   than  the  Salpeter    value, at x=0.05
\cite{96REI_EA}.   At  higher masses, the  IMF  is much more uncertain
(see Scalo (1998),  Haywood et al. (1997b) for  a review).   We
consider that  values  between x=1 and  2 are  reasonable, and  values
between 2 and 3 are not unrealistic.

\begin{figure}
\centering                  
\begin{center}
\epsfig{file=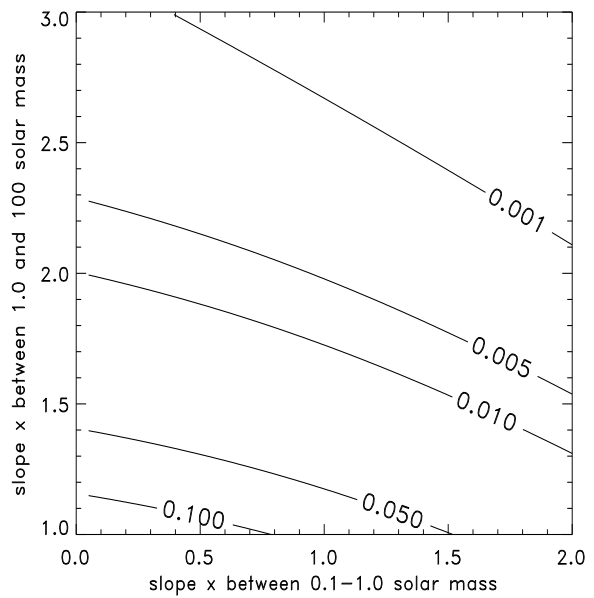,width=8.5cm,height=8.5cm}
\end{center}  
\caption{\em Yield weighted by the IMF as a function of the IMF
slope for low mass stars ($<1$M$_{\odot}$) and high mass stars ($>$1M$_{\odot}$).
The yield is calculated assuming stellar yields at solar metallicity, with specifications
mentioned in the text.
} 
\label{yieldfig1}
\end{figure}   

Fig.~\ref{yieldfig1} shows that for values of the  high mass IMF index
less than $\approx$ 1.9,  the yield is larger  than 0.01. For Salpeter
IMF index, values of the yield as high as 0.05 are possible. 

Assuming  that the  present ratio S$_{gas}$(t)/S$_{gas}$(0)  is around
0.1--0.3 and that the abundance of  the interstellar medium is of the
order 0.02--0.03, this value is in the range of possible yields.

\subsubsection[]{Surface and volume densities}\label{scaleheight}

An important issue  when discussing the distribution  of metallicities
is  the surface densities  of  the stellar and   gas components in the
solar neighbourhood.  According  to \shortcite{97JAH_EA}, the local  stellar
density is 3.9$\times$10$^{-2}$M$_{\odot}$.pc$^{-3}$.  The  projection
to   distance  outside  the  galactic  plane   can  be   made assuming
exponential density for  the thin disc and the  thick  disc.  Assuming
the thick disc  is responsible for 2 per cent  of the local stellar mass, and
has a scale  height of 1400~pc  \shortcite{93REI_EA},  the thin disc  with
325~pc, then the relative   amount of thick disc  is  of the order  of
8--9 per cent.  The  total surface  density  of visible stars  (i.e no stellar
remnants) is of the order  of  27 M$_{\odot}$.pc$^{-2}$.  Taking  into
account the gas surface density and allowing for some stellar remnant,
the     total    surface  density  is      of   the    order of  40--50
M$_{\odot}$.pc$^{-2}$.    Assuming    disc  characteristics as  those
proposed by \shortcite{97HAY_EA2}   and \shortcite{96ROB_EA}, one  finds that  the
thick disc represents 15 per cent of the local stellar surface density.

A correlated problem  is the correction  one has to  apply in order to
convert volume densities  to surface densities,  and vice versa.  This
is a difficult  problem,  because the  scale heights of  the (thin and
thick) discs are a function  of age, which  is not a quantity that can
be derived for  stars in the  sample. Various attempts have been made,
see in particular Sommer-Larsen (1991) and Wyse \& Gilmore (1995).   
Wyse \& Gilmore (1995) estimated  a correction function of the metallicity.    
One   problem   with  their   solution   is   that
they use   the same correction  for   stars in the interval
[Fe/H]=[-0.3,+0.3].    However,  even though this   may  seem a rather
narrow range, there is a significant age trend over this interval (see
the  age-metallicity relation  below).   It
implies that such stars may have quite different ages, 
hence quite  different correction  should  be applied.   We  choose to
`correct' the predicted  distribution instead (that is, we convert model surface density
to volume density), since metallicity is an
explicit function of age in the model. 
This solution has the advantage
of being  consistent with  the  fact  that the  vertical  velocity
dispersion is also a known function of age.

Quantitative  estimates of the  corrections as a  function of age have
been    derived     using  the    oscillation     period     given  in
Wyse \& Gilmore (1995).    We combine this  relative oscillation period
with the age-vertical velocity dispersion relation given in G\'omez et
al.  (2001).  The  combination of these two  factors gives  a relative
correction of the order of  13 for 'thick disc'  stars, and 5 for the
oldest disc  stars. Note that these  values are much larger than that
applied by Wyse \& Gilmore (1995), and partly  explains why we succeed
in giving a  reasonable fit to  the data.  Note  also that these values
are compatible with  scale  height variations within  the (thick
and thin) disc.

\subsection{Local constraints on the Simple Closed-Box model with no IRA}

Since the previous section shows the uncertainty introduced by converting
[Fe/H] to [O/H], we preferentially work with the iron distribution.
This means that we have to drop the IRA and calculate numerically the
metallicity distribution.

\subsubsection{The model}

The   main   characteristics  of the   model    are   given in   Table
\ref{SCBparam}.   The stellar yields are  used as described in section
\ref{parameters} for massive and intermediate mass stars.  The stellar
lifetimes  come from   \shortcite{97POL_EA},  and   are dependent  on
metallicity.   The onset   of  type Ia  supernovae    happens when the
metallicity reaches [Fe/H]=-1.0.  
The yields for the different species
produced by type Ia SN are from Nomoto et al. (1984) (iron $\approx$ 0.6 M$_{\odot}$/event)
and the rate of SNIa is assumed to be proportional to the number of SNII, with
SNII/SNIa=8.5.
Note that we don't use a Schmidt law
type SFR. While there are some evidences that the SFR of massive stars
may be proportional to some  power of the gas  density, there are few
evidences  that this can be extrapolated  to low and intermediate mass
stars. Since determinations to date are compatible with a constant SFR
history for the disc of the  Milky Way, we use this
simple prescription\footnote{It is expected, and perhaps has been
demonstrated that to some level, the SFR has not remained constant,
on time scale $<$1-2Gyrs. This is unimportant for the point considered here.
What we mean is that there is no demonstration that the SFR has decreased 
on increased systematically over 10-12Gyr.}.

\subsubsection{Metallicity distribution}

The delayed ejection  of important quantities  of  gas from long-lived
stars could  enhanced  the dilution   of  metals in  the  interstellar
medium,  and permit the production  of more stars at intermediate (e.g
solar) abundance.  In our tests however, this effect proved to be very
minor, for the reason   that the release  of  gas  is spread on   long
time-scale by  the  important   variation  of stellar   lifetime  with
abundance and mass; therefore, there is no sudden  release of gas, and
the  dilution  of metals in the  interstellar  medium is smoothed over
long time scales. 

\begin{figure}
\centering                  
\begin{center}
\epsfig{file=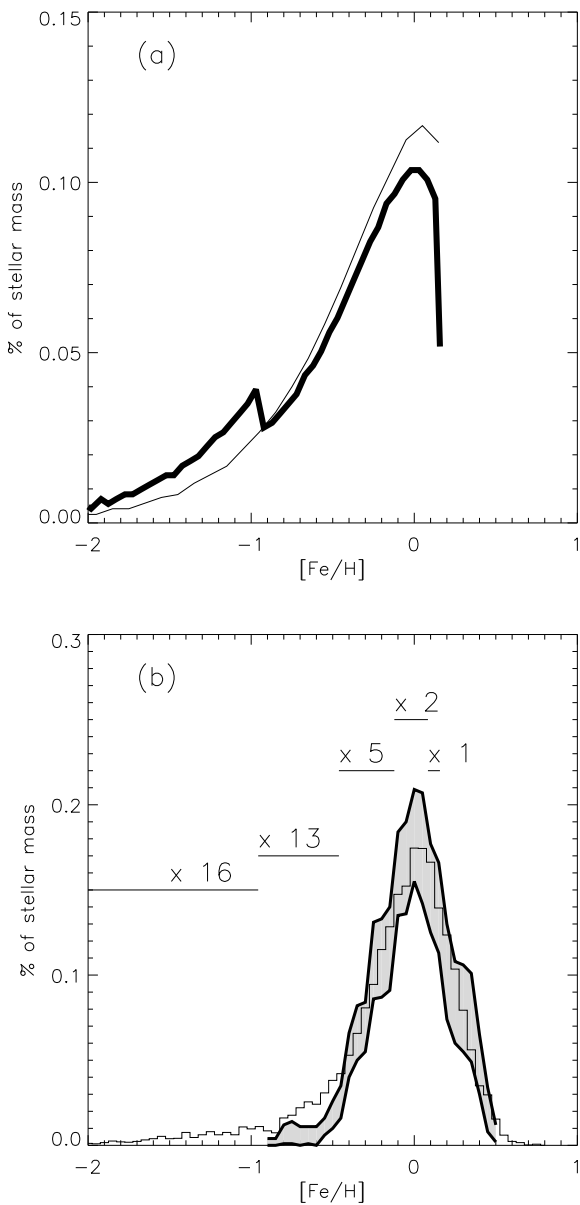,width=8.5cm}
\end{center}  
\caption{\em Plot (a) shows two Closed-Box models, one with IRA and yield p=0.02 (thin line) and no IRA (thick line), as described in the text. 
Plot (b) shows the SCB model (no IRA), converted to volume density as described in section 5.2.2, to which we have added a dispersion of 0.1~dex to account for metallicity measurement dispersion in the observed distribution.
} 
\label{closeboxnoira}
\end{figure}

The only marked effect of the SCB model with no IRA is the more rapid
metallicity increase at the onset of SNIa at [Fe/H]=-1. This translates
in  the metallicity distribution into a   shift of the distribution to
higher  metallicities,   as   can    be  seen   at   [Fe/H]$\approx$-1
(Fig.\ref{closeboxnoira}a). The SCB   with no  IRA  doesn't  seems  to
generate  more   solar metallicity   stars. Note  that   this somewhat
contradicts the  findings of Scully et al. (1997).  However, they use
very different  prescriptions    (stellar  lifetimes independent    of
metallicity, IMF index=1.7 over the  whole mass range, which they take
to be 0.4$\le$ m/M$_{\odot}$ $\leq$ 100).

\begin{table}
 \centering
  \caption{Characteristics of the SCB model of Fig.15.}
  \begin{tabular}{@{}llr@{}}
Surface density &   40M$_{\odot}$.pc$^{-2}$  \\
Initial metallicity & Fe/H=0.\\ 
SFR             &   Constant 3.5M$_{\odot}$.pc$^{-2}$.Gyr$^{-1}$ \\
IMF             &   x=0.05 m$<$ 1M$_{\odot}$, x=1.7 m$>$ 1M$_{\odot}$ \\
Final gas density & 8M$_{\odot}$.pc$^{-2}$ \\
Age of the model & 14 Gyr \\
\end{tabular}
\label{SCBparam}
\end{table}

The total duration of the  model is 14  Gyr. In the model, we identify
three  phases    corresponding   (not   necessarily  in    a  univocal
correspondance) more or less to the galactic stellar populations.  The
first phase is defined by [Fe/H]$<$-1.0~dex.  We do not try to ascribe
a particular  stellar population (halo or thick   disc) to this phase,
and the scale height  correction attributed to this  metallicity range
is   arbitrary.  In other   words, we don't include  this  part of the
distribution  in our discussion of the  SCB  model.  This conservative
position   is justified    to some   extent   by  (1) the   fact that  the
characteristics of the metal weak thick disc (or flattened halo ?) are
still essentially unknown (2) the limited volume of our sample is not well suited
for studying an intrinsically rare population.
In the model, the duration of this phase  is 3 Gyr.  The
thick  disc phase  starts at [Fe/H]=-1.0,  and  may  last 2 to  3 Gyr,
ending when  the  metallicity reaches  -0.58 or  -0.45.  Then the thin
disc phase  may be identified  with the remaining  evolution, with the
metallicity rising from -0.58 or -0.48 to [Fe/H]=+0.17. 
The disc as a whole is characterised by having around 11 Gyr, and a metallicity
that evolves from [Fe/H]=-1 to [Fe/H]=+0.17.

Figure \ref{closeboxnoira}(b)  shows  the metallicity  distribution of
the SCB model with no  IRA after the disc thickening has been accounted for. A gaussian noise of 
0.15~dex has been added
to account for  metallicity    dispersion at  all ages,  plus 0.1~dex
for simulating individual  measurement error.    The figure shows  a   quite
reasonable agreement between the    model and the   observations.

\subsubsection{The age-metallicity relation}

\begin{figure}
\centering                  
\begin{center}
\epsfig{file=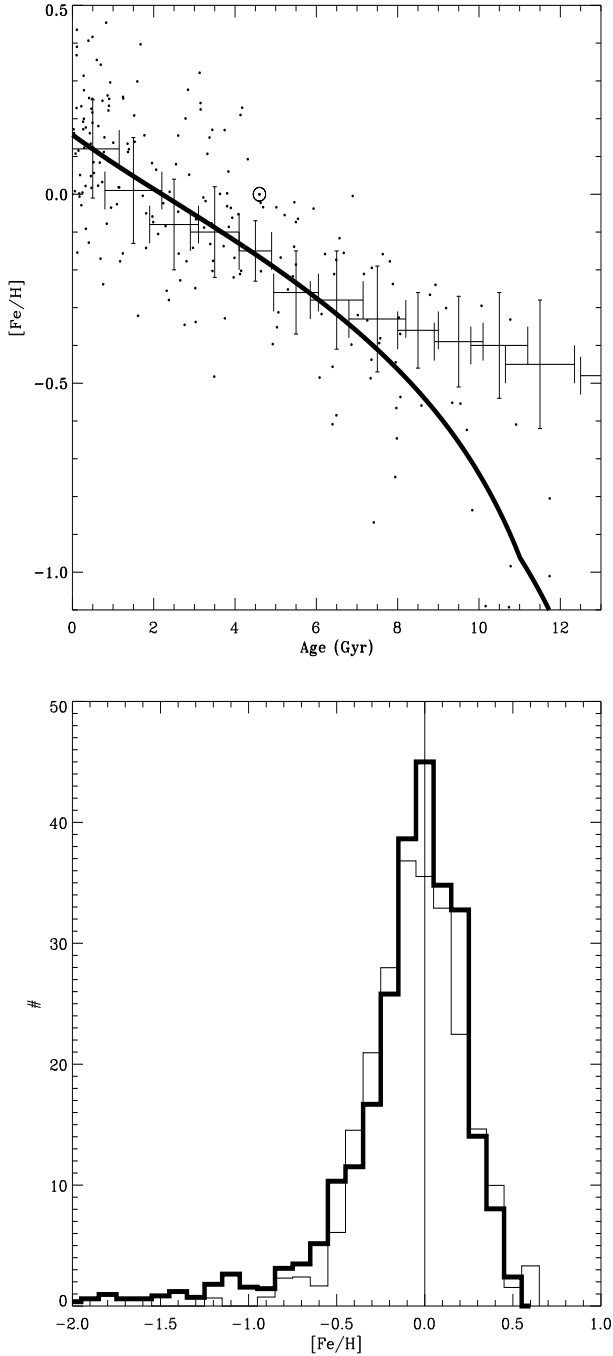,width=8.5cm,height=18.5cm}
\end{center}  
\caption{\em (a) The mean age-metallicity relation given by the model (thick curve).
The dots are calculated according to the same model, but assuming 10 per cent dispersion on the IMF, SFR,
and SN1 rate, plus 0.1~dex `measurement error' in [Fe/H] and age. This simulated
`local' distribution has been weighted according to the disc thickening described in section 5.2.2.
The points with error bars are the observed age-metallicity relation according to  \protect\shortcite{00ROC_EA2}.
Note that the two points corresponding to ages 11.5 and 13.5 Gyr are only upper estimates. 
(b) Shows the histogram of metallicities for these points (thick line).
The histogram of Fig. 6 is also shown (thin line).
}
\label{agefeh}
\end{figure}

On Fig.~\ref{agefeh}(a), we display the age-metallicity relation (AMR) of the model, together with the one
derived by \shortcite{00ROC_EA2}. The plot shows a reasonable agreement between 
the two over a large age range. Note that the two points corresponding to very old stars 
are upper limits according to \shortcite{00ROC_EA2}.

The age-metallicity relation has been  the subject of numerous studies
since  the work of \shortcite{93EDV_EA2}.  Because the  cosmic dispersion in
the  relation  is  viewed  as an   important  constraint  for chemical
evolution models,   it    has focused a    number of    comments  (see
\shortcite{00ROC_EA2} for a review). 
The  dispersion  in  the AMR  measured   by  \shortcite{93EDV_EA2} is  about
0.2~dex,   larger   than the   one   given  by  more   recent studies.
\shortcite{93EDV_EA2} cautioned  however that their  sample has  been chosen
for representativeness    of all metallicities,   which  doesn't  mean
completeness at all metallicities.   This makes their sample ill-suited
for  a correct measurement  of   the real dispersion.   Although  they
provide corrections for the   bias,  their procedure is   necessarily
uncertain.   It    is  not   surprising    that   \shortcite{00GAR_EA}   and
\shortcite{00ROC_EA2}   find a smaller    dispersion,    of the order     of
0.10--0.15~dex. A  consequence   on the AMR  is that    the correlation
between age  and metallicity  is  tighter than has been  thought after
\shortcite{93EDV_EA2}.    This view   confirms   that the distinction   made
(section   5.2)   in the scale   height  corrections   for  stars with
metallicity varying between -0.5 and +0.2 is justified. 

There are three main processes by which the dispersion may be explained: inhomogeneity
in the interstellar medium \shortcite{93MAL_EA}, diffusion of stellar orbits 
\cite{87GRE,93FRA_EA,96WIE_EA}, and sporadic infall \shortcite{97HOE_EA1}.
While it is almost certain that the first two play a role, even if minor,
infall has the capability to explain large dispersion in the age-metallicity relation.

Whatever the processes that lead to variations of the conditions of the star birth, it is worth noting
that a 10 per cent  dispersion on the IMF, SFR and SN1 rate is sufficient 
to account for the   observed dispersion. This is illustrated in Fig.\ref{agefeh}a), where
each point has been simulated as the result of SCB models with slightly different 
properties simulated at random from the model of table \ref{SCBparam} allowing for
10 percent dispersion in the IMF slope (for m$>$ 1M$_{\odot}$), SNIa rate and
SFR. A weighting according to the disc thickening described in section \ref{scaleheight}
has been applied.
Fig.~\ref{agefeh}b shows the metallicity histogram for these simulated points, compared to 
the metallicity distribution of Fig. \ref{HistFeHcomp.ps}.

\subsubsection{Abundance ratios}

Our aim here is to demonstrate further how a model with simple prescriptions gives a satisfactory fit to
complementary local constraints such as abundance ratios.

\begin{figure}
\centering                  
\begin{center}
\epsfig{file=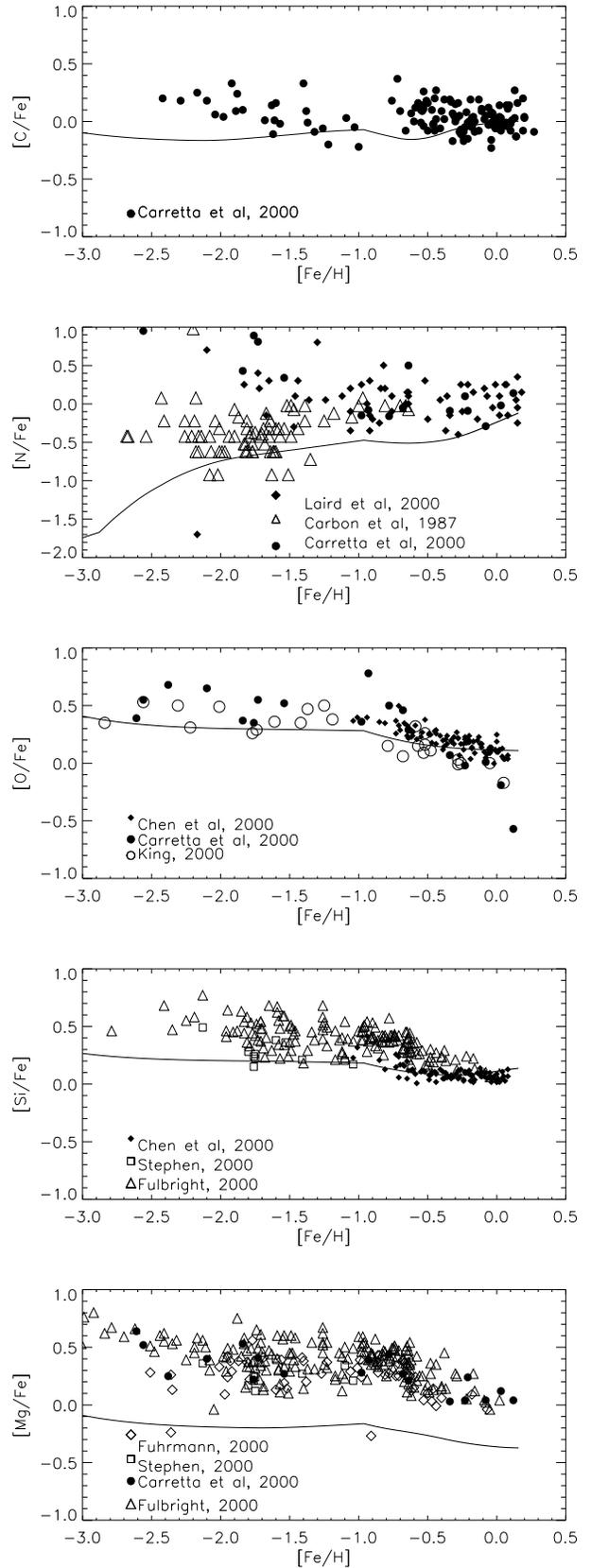,width=8.5cm}
\end{center}  
\caption{\em Abundance ratios for Carbon, Nitrogen, Oxygen,  Silicon and Magnesium. \protect\nocite{99STE_EA}
} 
\label{a_ratios}
\end{figure}

Fig.\ref{a_ratios} displays these ratios for CNO and Si, Mg elements.
The model has obvious failures (N/Fe, Mg/Fe) and reasonable success (O/Fe, C/Fe), 
in comparable proportion to more sophisticated models \cite{95TIM_EA,98POR_EA,98GOS_EA}.
Possible reasons for the failure of N/Fe, Mg/Fe are discussed in \shortcite{98GOS_EA}, and 
we don't replicate their discussion.

\section{Discussion}

A reanalysis of the solar neighbourhood metallicity distribution has
brought the following results :

(1) The  metallicity distribution  of  long-lived dwarfs is centred  on
[Fe/H]$\approx$0, not [-0.3,-0.1].   
When considering a sample of stars with no bias on the age, 
between 40 and 50 percent of stars have metallicity higher than [Fe/H]=0.
Most previous studies have used samples biased against stars with solar
metallicity or higher. This result reconciles the age-metallicity relation with
the dwarf metallicity distribution.

(2) The percentage of mid-metal-poor stars ([Fe/H]$<$-0.5) is in agreement
with estimates of the thick disc density from remote star counts and is
situated between 2-4 per cent. The thin disc is not an important contributor to
stars with [Fe/H]$<$-0.5.

(3) 40 per cent of these long-lived stars in our sample are X-ray emitters, which suggest
that a large part of the sample may consist of young stars (age $<$2 Gyr).
This supports the idea that the SFR is not a decreasing function of time.

(4) We have evaluated the distribution of stellar {\it mass} with metallicity.
Due to selection criteria of observed samples, stars of varying metallicities 
are sampled on unequal mass intervals.
We show that if this effect is left uncorrected, a second important bias against 
metal-rich stellar mass is introduced.

(5) We argue that a reasonable fit  to the [Fe/H] distribution can be obtained 
with a SCB model with no instantaneous recycling 
approximation. Satisfactory fits are also obtained for the age-metallicity
relation and abundance ratios.
These results suggest that the solar neighbourhood metallicity distribution 
contains few (or no) indications that long time-scale infall may have play a major
role in the building of the galactic disc.

In their study, Wyse \& Gilmore (1995)  commented that  `even were the  Simple
Box model to fit some dataset, its  inherent implausibility means that
it is more likely that several compensating effects had generated this
agreement by chance'.  How much unrealistic is the SCB model presented
here  ? Our model incorporates  all  ingredients
usually found  in chemical evolution models  of the  Galaxy, (i.e metallicity-dependent
stellar yields and lifetimes, empirical IMF and SFR), and satisfies local constraints.   
The only marked  feature absent in
our model is infall. While infall is  prevalent in chemical evolution
models, it is merely viewed as an  additional free parameter to control
the dilution  of metals in the  interstellar  medium. In this respect,
the SCB model is as much realistic as other models, but  it is
simpler. However,  the remark of Wyse \& Gilmore (1995)  still holds, and the
assumptions of the SCB models should be  looked at with even more scrutiny.

The SCB model is basically a  description of the chemical evolution of
the {\it  mean}  parameters  of the  galactic disc.  What  we  seek in
fitting the  observed distribution  presented  here is to know  if the
evolutionary  path  followed by  the  SCB model is the  correct one.
This approach has a chance of being sound if the assumptions of
the SCB models do not contradict observed features.
We shortly  review the main  assumptions of the SCB model.\\ 

(1)  Initial metallicity Fe/H=0.  
It  was  not necessary   to    adopt a  non-zero  initial
metallicity to account for the  observed metallicity distribution.  If
correct, this implies that  stars at arbitrarily low metallicities  with
disc  kinematics must exist. Morrison et al. (1990) found that thick  disc
stars  possibly reach  as  low  metallicities  as [Fe/H]=-1.60,  while Martin \& Morrison (1998)
have been able  to  show that objects with  thick
disc kinematics can reach [Fe/H]=-2.05. It  remains to be demonstrated
that thick disc metallicities can  reach even lower values. Note  that
at the  same time, \shortcite{98FUH} showed  that  the halo and thick
disc may be contemporary.

(2)    Constant  IMF. There is   to  date  no clear-cut  evidence for
systematic  variations in the IMF.  The  case  for a universal IMF has
been         considered  by           different     authors   (Kroupa,
astro-ph/0009005; Gilmore (1999))  and  is  favored  at  low masses
($<$1M$_{\odot}$).  The diagnostic for a universal IMF is much less clear
for masses   of  concern in    nucleosynthesis   \shortcite{98SCA}.
Variable IMF have been envisaged in  the context of chemical evolution
but has  received  relatively poor support  (e.g Martinelli \& Matteucci (2000)
and Chiappini et al. (2000) for recent attempts).

(3) Homogeneous interstellar  medium.  As mentioned  above, this is  a
minor assumption of the   SCB model, introduced  for analytical
tractability. Inhomogenity can be introduced in the model while keeping
with the SCB model as the main evolutionary path for chemical evolution. 
For example, it may be conceived that the disc as a whole has
behaved  as  an ensemble of  sub-SCB   models with  slightly varying
properties.  This  would not seriously  affect the picture of  the SCB
model as the progressive enrichment of a gaseous disc initially free of
metals.

(4) Closed Box. This arguably  is  the most challenging hypothesis  of
the SCB  model. Note that what we have assumed with this hypothesis
is that the surface density of the model has remained constant, therefore
fixing the rate of dilution of metals. In   sophisticated  models where the   Closed-Box
hypothesis is not  assumed, infall is  essentially viewed as  a way to
regulate  the  metal  enrichment,  in  order to
provide a solution to  fitting the observed metallicity  distribution.
It works  essentially as an additional free  parameter. In view of the
work presented here,   infall looks unnecessary to
describe the  solar  vicinity metallicity distribution.  The local fossil signature
of infall is still to be found. 
We  note that  the   potentially most
serious  indication that infall  processes  may have played an  important
role  in the   building  of  the   disc lies   in the  age-metallicity
relation.

Infall provides the only mechanism to generate old metal-rich (solar-like) stars
at the solar radius. If it can be demonstrated that  old stars (age$>$8 Gyr) with
disc-like kinematics and solar-like abundance exist in significant proportion, 
then they may represent the signature that  the disc, or  part of the (old) disc has
formed  through an evolutionary path  that  differs from the SCB model
evolution.

\section*{Acknowledgments}

This study was largely inspired by 
papers by M. Grenon, and his repeated cautionary advice that most
local surveys are biased against metal-rich stars.
I thank the referee Bernard Pagel for his many helpful comments
that improved this paper. 
This work has made extensive use of the Simbad database 
operated at CDS, Strasbourg, France.It is based on data
from ESA astrometric satellite $Hipparcos$.
\bsp

\label{lastpage}

\begin{table*}
\small
 \begin{minipage}{140mm}
  \caption{The list of 348 stars. Selected stars are marked with a "*"}
  \begin{tabular}{@{}rrllrrrllr@{}}
 HIP & [Fe/H] & M$_v$ & $B-V$ & & HIP & [Fe/H] & M$_v$ & $B-V$ & \\ \hline
 100925 & -0.02 & 5.17& 0.72&    &  39342 &  0.14 & 5.99 & 0.87 &  * \\
 101997 & -0.60 & 5.53& 0.72& *  &  40118 & -0.39 & 5.14 & 0.68 &  * \\
 102264 & -0.37 & 5.19& 0.67& *  &  40693 &  0.03 & 5.47 & 0.75 &  * \\
 103859 &  0.11 & 6.25& 0.97& *  &  40774 & -0.34 & 6.16 & 0.90 &  * \\
 105038 &  0.02 & 6.87& 1.02& *  &  41926 & -0.43 & 5.95 & 0.78 &  * \\
 105152 &  0.15 & 6.81& 0.99& *  &  42074 &  0.04 & 5.63 & 0.79 &  * \\
 105905 & -0.37 & 6.82& 0.92& *  &  42499 & -0.21 & 6.29 & 0.83 &  * \\
 106696 & -0.36 & 6.29& 0.88& *  &  42808 & -0.21 & 6.33 & 0.92 &  * \\
 107022 &  0.02 & 5.34& 0.75& *  &  43587 &  0.35 & 5.47 & 0.87 &  * \\
 107625 &  0.09 & 6.75& 0.96& *  &  43726 &  0.11 & 4.84 & 0.66 &    \\
 108156 &  0.11 & 6.23& 0.91& *  &  45170 & -0.45 & 4.94 & 0.73 &  * \\
 109378 &  0.22 & 4.91& 0.77& *  &  46580 & -0.06 & 6.69 & 1.00 &  * \\
 109527 & -0.03 & 5.51& 0.81& *  &  46626 &  0.21 & 6.95 & 0.99 &  * \\
 110649 &  0.03 & 3.75& 0.67&    &  46816 &  0.60 & 6.50 & 0.93 &  * \\
 111888 & -0.14 & 6.70& 0.94& *  &  46843 & -0.14 & 5.76 & 0.78 &  * \\
 112190 &  0.31 & 6.44& 0.97& *  &  49366 & -0.10 & 6.34 & 0.89 &  * \\
 112870 & -0.36 & 6.68& 0.85& *  &  51271 &  0.17 & 7.01 & 1.04 &  * \\
 113421 &  0.39 & 4.68& 0.74& *  &  51819 &  0.06 & 5.68 & 0.82 &  * \\
 114416 &  0.01 & 7.16& 1.00& *  &  52462 & -0.09 & 6.05 & 0.87 &  * \\
 115331 &  0.01 & 5.66& 0.80& *  &  53486 &  0.09 & 6.12 & 0.92 &  * \\
 115445 & -0.12 & 6.35& 0.88& *  &  54426 &  0.11 & 6.56 & 0.94 &  * \\
 116085 &  0.20 & 5.63& 0.84& *  &  54704 & -0.02 & 5.37 & 0.76 &  * \\
 116745 &  0.17 & 6.81& 0.99& *  &  55210 & -0.22 & 5.57 & 0.73 &  * \\
 116763 & -0.29 & 5.82& 0.80& *  &  56452 & -0.34 & 6.07 & 0.81 &  * \\
 118008 &  0.01 & 6.51& 0.97& *  &  56829 & -0.10 & 6.78 & 0.98 &  * \\
  10138 & -0.25 & 5.93& 0.81& *  &  56997 & -0.15 & 5.43 & 0.72 &  * \\
  10798 & -0.80 & 5.83& 0.72& *  &  57443 & -0.34 & 5.06 & 0.66 &  * \\
  12158 &  0.12 & 6.17& 0.94& *  &  57507 & -0.36 & 5.23 & 0.68 &  * \\
  13402 & -0.09 & 5.96& 0.86& *  &  58451 &  0.30 & 6.35 & 0.97 &  * \\
  13976 &  0.24 & 6.11& 0.93& *  &  58576 &  0.33 & 5.00 & 0.76 &  * \\
  14150 &  0.16 & 5.04& 0.70&    &  59280 &  0.16 & 5.53 & 0.79 &  * \\
  15099 &  0.10 & 6.09& 0.86& *  &  61291 & -0.18 & 6.09 & 0.84 &  * \\
  15457 &  0.06 & 5.02& 0.68&    &  61451 &  0.10 & 6.19 & 1.02 &  * \\
  15510 & -0.35 & 5.35& 0.71& *  &  61946 &  0.05 & 6.44 & 0.95 &  * \\
  16537 & -0.14 & 6.19& 0.88& *  &  62229 &  0.00 & 6.30 & 0.94 &  * \\
  17420 &  0.10 & 6.36& 0.93& *  &  62523 &  0.18 & 5.12 & 0.70 &    \\
  17439 & -0.01 & 5.94& 0.87& *  &  64690 & -0.06 & 5.16 & 0.71 &  * \\
  18324 & -0.05 & 6.21& 0.83& *  &  64924 &  0.01 & 5.09 & 0.71 &  * \\
  18915 & -1.60 & 7.17& 0.86& *  &  65530 &  0.22 & 4.86 & 0.74 &  * \\
  19422 &  0.14 & 6.38& 0.95& *  &  66147 &  0.28 & 6.65 & 1.03 &  * \\
  21988 &  0.15 & 6.26& 0.91& *  &  66765 & -0.03 & 5.97 & 0.86 &  * \\
  22122 & -0.02 & 6.03& 0.88& *  &  67620 & -0.10 & 4.94 & 0.70 &  * \\
  24874 & -0.11 & 6.83& 1.00& *  &  67655 & -0.93 & 5.98 & 0.66 &  * \\
  25421 &  0.25 & 6.46& 0.95& *  &  69357 &  0.00 & 6.11 & 0.87 &  * \\
  25544 & -0.23 & 5.53& 0.75& *  &  69414 & -0.07 & 5.31 & 0.73 &  * \\
  26779 & -0.01 & 5.79& 0.84& *  &  69972 &  0.39 & 6.29 & 1.02 &  * \\
  27207 &  0.15 & 5.79& 0.83& *  &  70016 &  0.09 & 5.99 & 0.87 &  * \\
  27887 &  0.07 & 6.62& 0.94& *  &  70950 &  0.06 & 6.99 & 1.03 &  * \\
  28954 & -0.01 & 5.83& 0.81& *  &  71181 &  0.16 & 6.61 & 1.00 &  * \\
  29271 &  0.08 & 5.05& 0.71& *  &  71395 &  0.08 & 6.39 & 0.97 &  * \\
  29525 & -0.15 & 5.15& 0.66&    &  72312 & -0.05 & 6.31 & 0.89 &  * \\
  29568 & -0.16 & 5.26& 0.71& *  &  72688 &  0.04 & 6.67 & 1.04 &  * \\
  32010 &  0.17 & 6.89& 1.02& *  &  73005 & -0.22 & 5.88 & 0.79 &  * \\
  33690 &  0.12 & 5.50& 0.79& *  &  74702 & -0.06 & 5.96 & 0.83 &  * \\
  33852 &  0.36 & 6.44& 0.99& *  &  75253 &  0.42 & 6.27 & 0.97 &  * \\
  34414 & -0.15 & 6.59& 0.91& *  &  75722 &  0.15 & 5.97 & 0.87 &  * \\
  36210 & -0.05 & 4.96& 0.69& *  &  75829 & -0.11 & 5.62 & 0.80 &  * \\
  38228 & -0.12 & 5.21& 0.68& *  &  76375 &  0.48 & 5.91 & 0.95 &  * \\
  38784 & -0.20 & 5.40& 0.72& *  &  77358 &  0.06 & 5.09 & 0.71 &  * \\
  39064 &  0.25 & 5.89& 0.83& *  &  77408 & -0.23 & 5.77 & 0.80 &  * \\
  39157 & -0.69 & 5.87& 0.72& *  &  78775 & -0.50 & 5.87 & 0.73 &  * \\
\hline
\end{tabular}
\label{tabledes350etoiles}
\end{minipage}
\end{table*}
\pagebreak
\addtocounter{table}{-1}
\begin{table*}
\small
 \begin{minipage}{140mm}
  \caption{(continued)}
  \begin{tabular}{@{}rrllrrrllr@{}}
  HIP & [Fe/H] & M$_v$ & $B-V$ &  & HIP & [Fe/H] & M$_v$ & $B-V$ & \\ \hline
  78913 & -0.09 & 6.74& 0.96& *  & 105184 & -0.06 & 4.87 & 0.80 &    \\
  79190 & -0.35 & 6.32& 0.86& *  & 105858 & -0.70 & 4.39 & 0.77 &    \\
  79248 &  0.43 & 5.35& 0.88& *  & 107350 & -0.04 & 4.64 & 0.75 &    \\
  79537 & -1.16 & 6.84& 0.81& *  & 107649 &  0.03 & 4.60 & 0.60 &    \\
  80366 & -0.09 & 6.73& 0.95& *  & 109422 &  0.09 & 3.58 & 0.69 &    \\
  81813 & -0.13 & 5.63& 0.77& *  & 109821 &  0.05 & 4.51 & 0.61 &    \\
  82588 & -0.18 & 5.49& 0.75& *  & 110109 & -0.27 & 4.69 & 0.51 &  * \\
  83389 & -0.27 & 5.48& 0.73& *  & 112117 & -0.15 & 4.13 & 0.59 &    \\
  83541 &  0.30 & 5.29& 0.81& *  & 113357 &  0.33 & 4.52 & 0.62 &    \\
  83990 & -0.22 & 6.71& 0.89& *  & 113829 & -0.01 & 4.72 & 0.64 &    \\
  85042 &  0.04 & 4.84& 0.68&    & 114886 & -0.16 & 6.15 & 0.49 &  * \\
  85235 & -0.44 & 5.90& 0.76& *  & 114924 & -0.10 & 4.04 & 0.59 &    \\
  86796 &  0.35 & 4.21& 0.69&    & 114948 & -0.14 & 4.07 & 0.60 &    \\
  87579 & -0.19 & 6.52& 0.94& *  & 115147 & -0.48 & 6.05 & 0.49 &  * \\
  88972 &  0.07 & 6.17& 0.88& *  & 116416 & -0.10 & 6.05 & 0.65 &  * \\
  90790 & -0.21 & 6.20& 0.86& *  & 116613 &  0.08 & 4.76 & 0.61 &    \\
  91438 & -0.32 & 5.29& 0.67& *  & 117712 & -0.29 & 6.19 & 0.58 &  * \\
  92858 & -0.17 & 6.08& 0.86& *  & 118162 &  0.15 & 4.80 & 0.67 &    \\
  92919 & -0.70 & 6.41& 0.91& *  &  12114 & -0.04 & 6.50 & 0.62 &  * \\
  93858 &  0.04 & 4.98& 0.70& *  &  12444 &  0.02 & 4.12 & 0.87 &    \\
  93966 &  0.18 & 4.48& 0.70&    &  12653 &  0.18 & 4.22 & 0.56 &    \\
  95319 &  0.12 & 5.42& 0.80& *  &  14286 & -0.22 & 4.84 & 0.52 &  * \\
  95447 &  0.40 & 4.25& 0.76&    &  14632 &  0.29 & 3.94 & 0.89 &    \\
  96085 &  0.06 & 6.26& 0.92& *  &  15131 & -0.40 & 4.82 & 0.85 &  * \\
  96100 & -0.14 & 5.87& 0.79& *  &  15330 & -0.21 & 5.11 & 0.67 &    \\
  96183 &  0.10 & 5.37& 0.75&    &  15371 & -0.22 & 4.83 & 0.98 &    \\
  96901 &  0.12 & 4.56& 0.66&    &  15442 & -0.18 & 5.09 & 0.69 &  * \\
  97944 &  0.38 & 5.43& 1.02&    &  16852 & -0.17 & 3.60 & 0.92 &    \\
  98130 &  0.05 & 7.00& 1.02& *  &  17147 & -0.82 & 4.75 & 0.52 &  * \\
  98677 & -0.29 & 5.73& 0.71& *  &  17378 &  0.12 & 3.74 & 0.56 &    \\
  98767 &  0.31 & 4.74& 0.75& *  &  18267 &  0.06 & 5.23 & 0.63 &    \\
  98792 & -0.28 & 6.32& 0.81& *  &  18859 &  0.09 & 3.96 & 0.60 &    \\
  98828 &  0.22 & 6.19& 0.92& *  &  19076 &  0.08 & 4.78 & 0.59 &    \\
  99240 &  0.36 & 4.62& 0.75&    &  19233 & -0.09 & 4.54 & 0.64 &    \\
  99711 &  0.13 & 6.27& 0.94& *  &  22263 &  0.12 & 4.87 & 0.60 &    \\
  99825 &  0.03 & 6.01& 0.88& *  &  22449 &  0.07 & 3.67 & 0.65 &    \\
   1031 & -0.20 & 5.68& 0.77& *  &  22451 & -0.16 & 6.22 & 0.57 &  * \\
   1292 & -0.03 & 5.36& 0.75& *  &  23437 & -0.54 & 5.28 & 0.55 &  * \\
   1499 &  0.14 & 4.61& 0.67&    &  23693 & -0.04 & 4.38 & 0.92 &    \\
   3093 &  0.26 & 5.65& 0.85& *  &  23835 & -0.08 & 3.91 & 0.72 &    \\
   3206 &  0.29 & 6.18& 0.94& *  &  24786 & -0.04 & 3.98 & 0.52 &    \\
   3535 &  0.56 & 6.32& 0.98& *  &  25278 & -0.04 & 4.17 & 0.62 &    \\
   3765 & -0.04 & 6.38& 0.89& *  &  26394 &  0.13 & 4.35 & 0.64 &    \\
   3850 & -0.37 & 5.78& 0.77& *  &  27072 & -0.00 & 3.83 & 0.63 &    \\
   3979 & -0.26 & 5.21& 0.66& *  &  27435 & -0.21 & 5.01 & 0.48 &  * \\
   4148 & -0.07 & 6.41& 0.94& *  &  27913 &  0.04 & 4.70 & 0.90 &    \\
   6379 & -0.10 & 6.02& 0.83& *  &  28267 & -0.05 & 5.16 & 0.64 &  * \\
   6917 & -0.25 & 5.91& 0.97& *  &  29432 &  0.01 & 5.03 & 0.53 &    \\
   7339 & -0.06 & 4.92& 0.69& *  &  29800 & -0.07 & 3.58 & 0.66 &    \\
   7576 & -0.05 & 5.80& 0.80& *  &  30314 &  0.02 & 4.68 & 0.57 &    \\
   7734 &  0.08 & 4.98& 0.69&    &  30503 &  0.14 & 4.65 & 0.54 &    \\
   8102 & -0.50 & 5.68& 0.73& *  &  30630 & -0.32 & 5.95 & 0.60 &  * \\
   8362 &  0.02 & 5.64& 0.80& *  &  32366 & -0.19 & 3.99 & 0.48 &    \\
   9269 &  0.20 & 5.18& 0.77& *  &  32423 & -0.12 & 6.81 & 0.64 &  * \\
    544 &  0.02 & 5.41& 0.75& *  &  32439 & -0.09 & 4.18 & 0.59 &    \\
 100017 & -0.13 & 4.69& 0.60&    &  32480 &  0.10 & 4.15 & 0.72 &    \\
 101345 &  0.12 & 3.74& 0.69&    &  33277 & -0.13 & 4.55 & 0.64 &    \\
 102040 & -0.16 & 4.82& 0.61&    &  33537 & -0.32 & 5.02 & 0.43 &  * \\
 103389 & -0.03 & 4.09& 0.51&    &  34017 & -0.11 & 4.53 & 0.61 &    \\
 103458 & -0.78 & 4.85& 0.59& *  &  34567 &  0.10 & 5.14 & 0.63 &    \\
 104436 & -0.42 & 5.06& 0.62& *  &  35136 & -0.33 & 4.41 & 0.94 &    \\
\hline
\end{tabular}
\end{minipage}
\end{table*}
\addtocounter{table}{-1}
\begin{table*}
\small
 \begin{minipage}{140mm}
  \caption{(continued)}
  \begin{tabular}{@{}rrllrrrllr@{}}
  HIP & [Fe/H] & M$_v$ & $B-V$ & &  HIP & [Fe/H] & M$_v$ & $B-V$ & \\ \hline
  36439 & -0.33 & 3.86 & 0.47 &     &  75181 & -0.26 & 4.83 & 0.64 &  * \\
  36515 & -0.16 & 4.97 & 0.64 &  *  &  75809 & -0.25 & 4.85 & 0.67 &  * \\
  36704 & -0.14 & 6.21 & 0.86 &  *  &  77052 &  0.12 & 5.03 & 0.68 &    \\
  37349 & -0.14 & 6.42 & 0.89 &  *  &  77257 &  0.11 & 4.07 & 0.60 &    \\
  38908 & -0.23 & 4.54 & 0.57 &     &  77801 & -0.49 & 4.86 & 0.60 &  * \\
  40035 & -0.17 & 3.77 & 0.49 &     &  78072 & -0.23 & 3.62 & 0.48 &    \\
  40843 & -0.23 & 3.84 & 0.49 &     &  79578 &  0.31 & 4.85 & 0.65 &    \\
  41484 &  0.04 & 4.63 & 0.62 &     &  79672 &  0.21 & 4.76 & 0.65 &    \\
  42333 &  0.13 & 4.87 & 0.65 &     &  80337 &  0.13 & 4.82 & 0.62 &    \\
  42438 & -0.14 & 4.86 & 0.62 &     &  80686 &  0.00 & 4.49 & 0.56 &  * \\
  42697 & -0.11 & 6.36 & 0.90 &  *  &  81300 &  0.00 & 5.82 & 0.83 &  * \\
  43557 & -0.08 & 4.66 & 0.64 &     &  81520 & -0.44 & 5.33 & 0.62 &    \\
  43797 &  0.11 & 3.79 & 0.48 &     &  83006 &  0.07 & 6.71 & 1.00 &  * \\
  44075 & -0.88 & 4.16 & 0.52 &     &  83601 &  0.00 & 4.45 & 0.58 &  * \\
  44897 &  0.11 & 4.54 & 0.58 &     &  84862 & -0.35 & 4.59 & 0.62 &  * \\
  45333 & -0.05 & 3.72 & 0.61 &     &  85653 & -0.53 & 5.47 & 0.74 &  * \\
  45617 &  0.19 & 5.98 & 0.99 &  *  &  85810 &  0.22 & 4.65 & 0.64 &    \\
  47080 &  0.35 & 5.16 & 0.77 &  *  &  86736 & -0.16 & 3.64 & 0.47 &    \\
  47592 & -0.09 & 4.07 & 0.53 &     &  86974 &  0.31 & 3.80 & 0.75 &    \\
  48113 &  0.23 & 3.75 & 0.62 &     &  88348 &  0.34 & 5.31 & 0.80 &  * \\
  50075 & -0.06 & 4.60 & 0.59 &     &  88622 & -0.51 & 4.86 & 0.61 &  * \\
  50384 & -0.55 & 4.03 & 0.50 &     &  88694 & -0.14 & 4.74 & 0.62 &    \\
  50505 & -0.23 & 5.09 & 0.65 &  *  &  89474 & -0.01 & 4.52 & 0.64 &    \\
  50921 & -0.27 & 5.20 & 0.66 &  *  &  89805 &  0.09 & 4.37 & 0.58 &    \\
  51248 & -0.39 & 4.56 & 0.61 &  *  &  93185 & -0.24 & 4.95 & 0.61 &    \\
  51933 & -0.39 & 3.76 & 0.53 &     &  96395 & -0.03 & 4.81 & 0.64 &    \\
  52369 & -0.10 & 4.94 & 0.62 &     &  96895 &  0.16 & 4.32 & 0.64 &    \\
  53721 &  0.13 & 4.29 & 0.62 &     &  97675 &  0.05 & 3.68 & 0.56 &    \\
  57939 & -1.66 & 6.61 & 0.75 &  *  &  98470 & -0.13 & 4.05 & 0.50 &    \\
  61053 &  0.03 & 4.49 & 0.57 &     &  98505 & -0.19 & 6.25 & 0.93 &  * \\
  61317 & -0.21 & 4.63 & 0.59 &     &  98959 & -0.23 & 4.83 & 0.65 &  * \\
  62207 & -0.61 & 4.75 & 0.56 &  *  &  99137 &  0.02 & 4.43 & 0.53 &    \\
  63366 & -0.27 & 5.93 & 0.77 &  *  &  99461 & -0.28 & 6.41 & 0.87 &  * \\
  64394 &  0.14 & 4.42 & 0.57 &     &   1598 & -0.40 & 4.99 & 0.64 &  * \\
  64457 & -0.11 & 6.01 & 0.93 &  *  &   1599 & -0.12 & 4.56 & 0.58 &    \\
  64550 & -0.08 & 4.99 & 0.64 &     &   1803 &  0.19 & 4.84 & 0.66 &    \\
  64583 & -0.29 & 3.62 & 0.49 &     &   3497 & -0.24 & 4.85 & 0.65 &  * \\
  64792 &  0.04 & 3.92 & 0.58 &     &   3909 & -0.11 & 4.22 & 0.51 &    \\
  65515 & -0.02 & 5.59 & 0.80 &  *  &   5862 & -0.03 & 4.08 & 0.57 &    \\
  67275 &  0.25 & 3.54 & 0.51 &     &   5944 & -0.09 & 4.72 & 0.59 &    \\
  68030 & -0.24 & 4.24 & 0.52 &     &   7235 &  0.00 & 5.52 & 0.77 &  * \\
  69671 & -0.14 & 4.69 & 0.60 &     &   7978 & -0.03 & 4.32 & 0.55 &    \\
  69965 & -0.71 & 4.62 & 0.52 &  *  &   9829 & -0.20 & 5.06 & 0.66 &  * \\
  70319 & -0.36 & 5.02 & 0.64 &  *  &    910 & -0.59 & 3.51 & 0.49 &    \\
  70873 &  0.28 & 4.50 & 0.70 &     &    950 & -0.08 & 3.55 & 0.46 &    \\
  71284 & -0.46 & 3.52 & 0.36 &     & 114622 &  0.20 & 6.50 & 1.00 &  * \\
  71743 &  0.16 & 5.38 & 0.71 &     &  12777 & -0.02 & 3.85 & 0.51 &    \\
  71855 & -0.02 & 5.19 & 0.71 &  *  &  19849 & -0.34 & 5.92 & 0.82 &  * \\
  72567 &  0.06 & 4.59 & 0.58 &     &  24813 & -0.03 & 4.18 & 0.63 &    \\
  73100 &  0.08 & 3.65 & 0.53 &     &  38939 & -0.17 & 7.12 & 1.04 &  * \\
  74273 &  0.13 & 4.38 & 0.62 &     &  49081 &  0.10 & 4.50 & 0.68 &    \\
  74537 &  0.13 & 5.39 & 0.76 &  *  &  54745 & -0.23 & 4.73 & 0.60 &    \\
\hline
\end{tabular}
\end{minipage}
\end{table*}

\pagebreak


\begin{thebibliography}{}

\bibitem[{{Alonso} et~al., 1996}]{96ALO_EA1}
{Alonso} A., {Arribas} S.,  {Martinez-Roger} C., 1996, Astronomy and
  Astrophysics Supplement Series, 117, 227

\bibitem[{{Axer}
  et~al., 1995}]{95AXE_EA}
{Axer} M., {Fuhrmann} K.,  {Gehren} T., 1995, Astron. Astrophys., 300, 751

\bibitem[{{Bertelli} et~al., 1994}]{94BER_EA}
{Bertelli} G., {Bressan} A., {Chiosi} C., {Fagotto} F.,  {Nasi} E., 1994,
  Astronomy and Astrophysics Supplement Series, 106, 275

\bibitem[{{Binney}
  et~al., 2000}]{00BIN_EA}
{Binney} J., {Dehnen} W.,  {Bertelli} G., 2000, Mon. Not. R. Astron. Soc., 318,
  658

\bibitem[{{Binney} \& {Merrifield}, 1998}]{98BIN_EA}
{Binney} J.,  {Merrifield} M., 1998, "Galactic astronomy".
\newblock Galactic astronomy / James Binney and Michael Merrifield. Princeton
  University Press. (Princeton series in astrophysics)

\bibitem[{{Binney} \& {Tremaine}, 1987}]{87BIN_EA}
{Binney} J.,  {Tremaine} S., 1987, "Galactic dynamics".
\newblock Princeton, NJ, Princeton University Press, 1987, 747 p.

\bibitem[{{Brinchmann} \& {Ellis}, 2000}]{00BRI_EA}
{Brinchmann} J.,  {Ellis} R.~S., 2000, Astrophys. J., Lett., 536, L77

\bibitem[{{Carretta}, {Gratton}, \& {Sneden}}{{Carretta}
  et~al., 2000}]{00CAR_EA}
{Carretta} E., {Gratton} R.~G.,  {Sneden} C., 2000, Astron. Astrophys., 356,
  238

\bibitem[{{Castro} et~al., 1997}]{97CAS_EA1}
{Castro} S., {Rich} R.~M., {Grenon} M., {Barbuy} B.,  {McCarthy} J.~K., 1997,
  Astron. J., 114, 376

\bibitem[{{Cayrel De Strobel} et~al., 1997}]{97CAY_EA}
{Cayrel De Strobel} G., {Soubiran} C., {Friel} E.~D., {Ralite} N.,  {Francois}
  P., 1997, Astronomy and Astrophysics Supplement Series, 124, 299, Provided by
  the NASA Astrophysics Data System

\bibitem[{{Chen} et~al., 2000}]{00CHE_EA}
{Chen} Y.~Q., {Nissen} P.~E., {Zhao} G., {Zhang} H.~W.,  {Benoni} T., 2000,
  Astron. Astrophys. Suppl. Ser., 141, 491

\bibitem[{{Chiappini} et~al., 1997}]{97CHI_EA}
{Chiappini} C., {Matteucci} F.,  {Gratton} R., 1997, Astrophys. J., 477, 765

\bibitem[{{Chiappini} et~al., 2000}]{00CHI_EA}
{Chiappini} C., {Matteucci} F.,  {Padoan} P., 2000, Astrophys. J., 528, 711

\bibitem[{{Clegg}
  et~al., 1981}]{81CLE_EA}
{Clegg} R.~E.~S., {Tomkin} J.,  {Lambert} D.~L., 1981, Astrophys. J., 250, 262

\bibitem[{{Edvardsson} et~al., 1993}]{93EDV_EA2}
{Edvardsson} B., {Andersen} J., {Gustafsson} B., {Lambert} D.~L., {Nissen}
  P.~E.,  {Tomkin} J., 1993, Astron. Astrophys., 275, 101

\bibitem[{{Favata}
  et~al., 1997}]{97FAV_EA2}
{Favata} F., {Micela} G.,  {Sciortino} S., 1997, Astron. Astrophys., 323, 809

\bibitem[{{Favata} et~al., 1997}]{97FAV_EA3}
{Favata} F., {Micela} G., {Sciortino} S.,  {Morale} F., 1997, Astron.
  Astrophys., 324, 998

\bibitem[{{Feltzing} \& {Gustafsson}, 1998}]{98FEL_EA}
{Feltzing} S.,  {Gustafsson} B., 1998, Astronomy and Astrophysics Supplement
  Series, 129, 237

\bibitem[{{Flynn} \& {Morell}, 1997}]{97FLY_EA}
{Flynn} C.,  {Morell} O., 1997, Mon. Not. R. Astron. Soc., 286, 617

\bibitem[{{Francois} \& {Matteucci}, 1993}]{93FRA_EA}
{Francois} P.,  {Matteucci} F., 1993, Astron. Astrophys., 280, 136

\bibitem[{{Fuhrmann}, 1998}]{98FUH}
{Fuhrmann} K., 1998, Astron. Astrophys., 338, 161

\bibitem[{{Garnett} \& {Kobulnicky}, 2000}]{00GAR_EA}
{Garnett} D.~R.,  {Kobulnicky} H.~A., 2000, Astrophys. J., 532, 1192

\bibitem[{{Gilmore}, 1999}]{99GIL}
{Gilmore} G., 1999, in The Identification of Dark Matter, p. 121

\bibitem[{{Gimenez} et~al., 1991}]{91GIM_EA}
{Gimenez} A., {Reglero} V., {de Castro} E.,  {Fernandez-Figueroa} M.~J., 1991,
  Astron. Astrophys., 248, 563

\bibitem[{{Gliese} \& {Jahrei{\ss}}, 1991}]{91GLI_EA}
{Gliese} W.,  {Jahrei{\ss}} H., 1991, Preliminary version of the third
  catalogue of nearby stars, Technical report

\bibitem[{{Goswami} \& {Prantzos}, 2000}]{98GOS_EA}
{Goswami} A.,  {Prantzos} N., 2000, Astron. Astrophys., 359, 191

\bibitem[{{Gratton} et~al., 2000}]{00GRA_EA}
{Gratton} R.~G., {Carretta} E., {Matteucci} F.,  {Sneden} C., 2000, Astron.
  Astrophys., 358, 671

\bibitem[{{Grenon}, 1978}]{78GRE}
{Grenon} M., 1978, Publications de l'Observatoire de Gen\`eve, 1

\bibitem[{{Grenon}, 1987}]{87GRE}
{Grenon} M., 1987, Journal of Astrophysics and Astronomy, 8, 123

\bibitem[{{Grenon}, 1990}]{90GRE2}
{Grenon} M., 1990, in Proceedings of the Fifth IAP Workshop, Astrophysical Ages
  and Dating Methods, June 26-30, 1989., p. 153

\bibitem[{{Guillout} et~al., 1998}]{98GUI_EA}
{Guillout} P., {Sterzik} M.~F., {Schmitt} J.~H. M.~M., {Motch} C.,
  {Neuhaeuser} R., 1998, Astron. Astrophys., 337, 113

\bibitem[{{Haywood}
  et~al., 1997}]{97HAY_EA2}
{Haywood} M., {Robin} A.~C.,  {Cr\'ez\'e} M., 1997, Astron. Astrophys., 320,
  440

\bibitem[{{Huensch} et~al., 1999}]{99HUE_EA}
{Huensch} M., {Schmitt} J.~H. M.~M., {Sterzik} M.~F.,  {Voges} W., 1999,
  Astron. Astrophys. Suppl. Ser., 135, 319

\bibitem[{{Iben} \& {Tutukov}, 1984}]{84IBE_EA}
{Iben} I.,  {Tutukov} A.~V., 1984, Astrophys. J., Suppl. Ser., 54, 335

\bibitem[{{Jahreiss} \& {Wielen}, 1997}]{97JAH_EA}
{Jahreiss} H.,  {Wielen} R., 1997, Proceedings of the ESA Symposium `Hipparcos
  - Venice '97', 13-16 May, Venice, Italy, ESA SP-402 (July 1997), p. 675-680,
  402, 675

\bibitem[{{Lebreton}, 2000}]{00LEB}
{Lebreton} Y., 2000, Ann. Rev. Astron. Astrophys., 38

\bibitem[{{Malinie} et~al., 1993}]{93MAL_EA}
{Malinie} G., {Hartmann} D.~H., {Clayton} D.~D.,  {Mathews} G.~J., 1993,
  Astrophys. J., 413, 633

\bibitem[{{Martin} \& {Morrison}, 1998}]{98MAR_EA}
{Martin} J.~C.,  {Morrison} H.~L., 1998, Astron. J., 116, 1724

\bibitem[{{Martinelli} \& {Matteucci}, 2000}]{00MAR_EA}
{Martinelli} A.,  {Matteucci} F., 2000, Astron. Astrophys., 353, 269

\bibitem[{Mermilliod et~al., 1997}]{97MER_EA1}
{Mermilliod} J.~., {Mermilliod} M.,  {Hauck} B., 1997, Astron. Astrophys.
  Suppl. Ser., 124, 349

\bibitem[{{Morale} et~al., 1996}]{96MOR_EA}
{Morale} F., {Micela} G., {Favata} F.,  {Sciortino} S., 1996, Astron.
  Astrophys. Suppl. Ser., 119, 403

\bibitem[{{Morrison}
  et~al., 1990}]{90MOR_EA}
{Morrison} H.~L., {Flynn} C.,  {Freeman} K.~C., 1990, Astron. J., 100, 1191

\bibitem[{Olsen, 1983}]{OLS}
{Olsen} E.~H., 1983, Astron. Astrophys. Suppl. Ser., 54, 55

\bibitem[{{Pagel}, 1989}]{89PAG1}
{Pagel} B.~E.~J., 1989, in Evolutionary Phenomena in Galaxies, p. 201

\bibitem[{{Pasquini} et~al., 1994}]{94PAS_EA}
{Pasquini} L., {Liu} Q.,  {Pallavicini} R., 1994, Astron. Astrophys., 287, 191

\bibitem[{{Perryman}
  et~al., 1998}]{98PER_EA}
{Perryman} M.~A.~C. et~al., 1998, Astron. Astrophys., 331, 81

\bibitem[{{Pilyugin} \& {Edmunds}, 1996}]{96PIL_EA}
{Pilyugin} L.~S.,  {Edmunds} M.~G., 1996, Astron. Astrophys., 313, 783

\bibitem[{{Pols} et~al., 1997}]{97POL_EA}
{Pols} O.~R., {Tout} C.~A., {Schroder} K., {Eggleton} P.~P.,  {Manners} J.,
  1997, Mon. Not. R. Astron. Soc., 289, 869

\bibitem[{{Portinari} \& {Chiosi}, 1999}]{99POR_EA}
{Portinari} L.,  {Chiosi} C., 1999, Astron. Astrophys., 350, 827

\bibitem[{{Portinari} et~al., 1998}]{98POR_EA}
{Portinari} L., {Chiosi} C.,  {Bressan} A., 1998, Astron. Astrophys., 334, 505

\bibitem[{{Prantzos} \& {Silk}, 1998}]{98PRA_EA}
{Prantzos} N.,  {Silk} J., 1998, Astrophys. J., 507, 229

\bibitem[{{Reid} et~al., 1996}]{96REI_EA}
{Reid} I.~N., {Hawley} S.~L.,  {Gizis} J.~E., 1996, Astron. J., 111, 2469

\bibitem[{{Reid} \& {Majewski}, 1993}]{93REI_EA}
{Reid} N.,  {Majewski} S.~R., 1993, Astrophys. J., 409, 635

\bibitem[{{Robin} et~al., 1996}]{96ROB_EA}
{Robin} A., {Haywood} M., {Cr\'ez\'e} M., {Ojha} D.,  {Bienayme} O., 1996,
  Astron. Astrophys., 305, 125

\bibitem[{{Rocha-Pinto} \& {Maciel}, 1996}]{96ROC_EA3}
{Rocha-Pinto} H.~J.,  {Maciel} W.~J., 1996, Mon. Not. R. Astron. Soc., 279, 447

\bibitem[{{Rocha-Pinto} \& {Maciel}, 1998}]{98ROC_EA}
{Rocha-Pinto} H.~J.,  {Maciel} W.~J., 1998, Astron. Astrophys., 339, 791

\bibitem[{{Rocha-Pinto} et~al., 2000}]{00ROC_EA2}
{Rocha-Pinto} H.~J., {Maciel} W.~J., {Scalo} J.,  {Flynn} C., 2000, Astron.
  Astrophys., 358, 850

\bibitem[{{Scalo}, 1998}]{98SCA}
{Scalo} J., 1998, in ASP Conf. Ser. 142: The Stellar Initial Mass Function
  (38th Herstmonceux Conference), p. 201

\bibitem[{{Scalo}, 1986}]{86SCA}
{Scalo} J.~M., 1986, Fundamentals of Cosmic Physics, 11, 1

\bibitem[{{Schuster} \& {Nissen}, 1989}]{89SCH_EA1}
{Schuster} W.~J.,  {Nissen} P.~E., 1989, Astron. Astrophys., 221, 65

\bibitem[{{Scully} et~al., 1997}]{97SCU_EA}
{Scully} S., {Casse} M., {Olive} K.~A.,  {Vangioni-Flam} E., 1997, Astrophys.
  J., 476, 521

\bibitem[{{Soderblom} et~al., 1991}]{91SOD_EA}
{Soderblom} D.~R., {Duncan} D.~K.,  {Johnson} D.~R.~H., 1991, apj, 375, 722

\bibitem[{{Sommer-Larsen}, 1991}]{91SO%
M1}
{Sommer-Larsen} J., 1991, Mon. Not. R. Astron. Soc., 249, 368

\bibitem[{{Stephens}, 1999}]{99STE_EA}
{Stephens} A., 1999, Astron. J., 117, 1771

\bibitem[{{Sterzik} \& {Schmitt}, 1997}]{97STE_EA}
{Sterzik} M.~F.,  {Schmitt} J.~H. M.~M., 1997, Astron. J., 114, 1673

\bibitem[{{Stetson} \& {Harris}, 1988}]{88STE_EA}
{Stetson} P.~B.,  {Harris} W.~E., 1988, Astron. J., 96, 909

\bibitem[{Th{\'e}venin \& Idiart, 1999}]{99THE_EA}
{Th{\'e}venin} F.,  {Idiart} T.~P., 1999, Astrophys. J., 521, 753

\bibitem[{{Timmes}
  et~al., 1995}]{95TIM_EA}
{Timmes} F.~X., {Woosley} S.~E.,  {Weaver} T.~A., 1995, Astrophys. J., Suppl.
  Ser., 98, 617

\bibitem[{{Tosi}, 1996}]{96TOS}
{Tosi} M., 1996, in ASP Conf. Ser. 98: From Stars to Galaxies: the Impact of
  Stellar Physics on Galaxy Evolution, p. 299

\bibitem[{{van den Hoek} \& {de Jong}, 1997}]{97HOE_EA1}
{van den Hoek} L.~B.,  {de Jong} T., 1997, Astron. Astrophys., 318, 231

\bibitem[{van den Hoek \& Groenewegen, 1997}]{97HOE_EA2}
{van den Hoek} L.~B.,  {Groenewegen} M.~A.~T., 1997, Astron. Astrophys., 123, 305

\bibitem[{{Wielen}
  et~al., 1996}]{96WIE_EA}
{Wielen} R., {Fuchs} B.,  {Dettbarn} C., 1996, Astron. Astrophys., 314, 438

\bibitem[{Wyse \& Gilmore, 1995}]{95WYS_EA}
{Wyse} R.~F.~G.,  {Gilmore} G., 1995, Astron. J., 110, 2771

\end{thebibliography}
\end{document}